\documentclass[aps,superscriptaddress,amsmath,twocolumn,amssymb,floatfix,showpacs,superscriptaddress,nofootinbib,longbibliography]{revtex4-1}
\usepackage{MnSymbol}
\usepackage{braket}
\usepackage[dvipsnames]{xcolor}
\usepackage{float}
\usepackage{subfigure}
\usepackage{tikz}
\usepackage[colorlinks=true,linktoc=page,linkcolor=OliveGreen,citecolor=purple,urlcolor=violet]{hyperref}

\mathchardef\mhyphen="2D % Define a "math hyphen"

\newcommand{\ie}{{i.e.,\,\,}}
\newcommand{\eg}{{e.g.,~}}

\newcommand\bea{\begin{eqnarray}}
\newcommand\eea{\end{eqnarray}}
\newcommand\beq{\begin{equation}}  
\newcommand\eeq{\end{equation}}

\newcommand{\non}{\nonumber}  
\usepackage[normalem]{ulem}
\definecolor{lime}{HTML}{A6CE39}
\usepackage{sidecap,tikz}
\DeclareRobustCommand{\orcidicon}{\hspace{-1.0mm}
	\begin{tikzpicture}
		\draw[lime, fill=lime] (0.0,0.0) 
		circle [radius=0.15] 
		node[white] {{\fontfamily{qag}\selectfont \tiny \,ID}};
		\draw[white, fill=white] (-0.0525,0.095) 
		circle [radius=0.007];
	\end{tikzpicture}
	\hspace{-3.0mm}
}
\foreach \x in {A, ..., Z}{\expandafter\xdef\csname orcid\x\endcsname{\noexpand\href{https://orcid.org/\csname orcidauthor\x\endcsname}{\noexpand\orcidicon}}
}

\AtBeginDocument{%
	\newwrite\bibnotes
	\def\bibnotesext{Notes.bib}
	\immediate\openout\bibnotes=\jobname\bibnotesext
	\immediate\write\bibnotes{@CONTROL{REVTEX41Control}}
	\immediate\write\bibnotes{@CONTROL{%
			apsrev41Control,author="08",editor="1",pages="1",title="1",year="1"}}
	\if@filesw
	\immediate\write\@auxout{\string\citation{apsrev41Control}}%
	\fi
}%

\begin{document}
%=============START of MAIN PAPER===============

\title{Chiral topological superconductivity in twisted bilayer and double bilayer graphene}  

\author{Kamalesh Bera\orcidA{}}
\email{kamalesh.bera@iopb.res.in}
\affiliation{Institute of Physics, Sachivalaya Marg, Bhubaneswar-751005, India}
\affiliation{Homi Bhabha National Institute, Training School Complex, Anushakti Nagar, Mumbai 400094, India}

\author{Tanay Nag\orcidB{}}
\email{tanay.nag@hyderabad.bits-pilani.ac.in}
\affiliation{Department of Physics, BITS Pilani-Hyderabad Campus, Telangana 500078, India}

\author{Arijit Saha\orcidC{}}
\email{arijit@iopb.res.in}
\affiliation{Institute of Physics, Sachivalaya Marg, Bhubaneswar-751005, India}
\affiliation{Homi Bhabha National Institute, Training School Complex, Anushakti Nagar, Mumbai 400094, India}

%--------------------------------------------------------
%--------------------------------------------------------
\begin{abstract}
We present a theoretical investigation of the emergence of chiral topological superconductivity in small-angle twisted bilayer graphene (tBLG) and twisted double bilayer graphene (tDBLG). Using the low-energy continuum model and incorporating spin-triplet $p_{x}+i p_{y}$ pairing in each graphene layer, we construct the effective models for both tBLG and tDBLG with superconductivity. By varying the chemical potential, superconducting order parameter, and twist angle, we explore the emergence of topological superconducting phases via the calculation of Chern numbers. Our phase diagrams for tBLG and tDBLG (both AB-AB and AB-BA stackings) reveal distinct topological transitions, which are consistently marked by bulk gap-closing points. To gain further insight, we analyze the evolution of Chern numbers by tracking the number and location of gap closings within the moir\'e Brillouin zone. Additionally, we illustrate representative squared amplitude of Bloch states corresponding to different topological phases. In the later part of our study, the effect of trigonal warping on the topological superconducting properties is also discussed. Beyond the quantitative results, our study highlights how the interplay between moir\'e band structure and unconventional pairing symmetries enriches the landscape of possible superconducting states in twisted graphene systems. The framework developed here may also be extended to other multilayer moir\'e materials, offering a route towards engineering exotic topological superconductivity with tunable parameters.
\end{abstract}
%--------------------------------------------------------
%--------------------------------------------------------

\maketitle

%======================================================
\section{Introduction}
%======================================================

%-------------- motivation --------------%
Chiral superconductors are a class of unconventional superconductors that spontaneously break time-reversal symmetry. Unlike conventional superconductors where Cooper pairs carry zero angular momentum chiral superconductors host Cooper pairs with finite angular momentum~\cite{Sigrist1991}. These systems can give rise to exotic quasiparticles such as Majorana fermions and provide an important platform for studying topological superconductivity~\cite{TI_TSC_RMP_review_Zhang,TSC_classification_Schnyder}. Majorana fermions, in turn, are predicted to play a crucial role in future technologies, particularly in the development of fault-tolerant quantum computation~\cite{TI_TSC_RMP_review_Zhang,Sato_2017,RMP_Nayak,TSC_classification_Schnyder,TSC_review_Kallin_2016,TSC_review_kitaev}. One of the best-known examples of a chiral topological phase is the superfluid $^3$He-A phase~\cite{He3_Leggett}. Over the past three decades, various experimental techniques have been employed to investigate potential chiral superconductors. While initial findings suggested compatibility with chiral superconductivity~\cite{Maeno1994,Saxena2000,Ghosh2021,Jiao2020,Hayes2021}, more recent studies indicate that materials such as UTe$_2$ and Sr$_2$RuO$_4$ possess single-component order parameter~\cite{Pustogow2019,Theuss2024}. However, very recently, signatures of chiral superconductivity have been more promisingly identified in rhombohedral graphene~\cite{Han2025}.

%Introduction to tBLG and tDBLG
In recent times, moir\'e systems have emerged as a highly tunable material platform, with the twist angle serving as a new control parameter. It was theoretically predicted that when two layers of graphene are rotated relative to each other, the kinetic energy becomes quenched at certain twist angles~\cite{Santos-Peres-tBLG,Shallcross-tBLG,MacDonald-tBLG,Koshino-tBLG}. The latter experimental discovery of unconventional superconductivity and Mott-like insulating phases in twisted bilayer graphene (tBLG) near the so-called magic angle~\cite{Cao2018-unconv_sc,Cao2018-corr_insulator} confirmed these theoretical predictions and established tBLG as a tunable platform for understanding and realizing various strongly correlated phases near the flat band~\cite{Oh2021-tBLG(xpt),Nuckolls2020-tBLG(xpt),Magnetism-tBLG,CIS-tBLG,MI+SC-tBLG,Wu2021-chernIns-expt,All_Magic_Angle-tBLG,Origin_of_Magic_Angle-tBLG,Heavy_fermion-tBLG}. Depending on the number of layers, stacking configurations, lattice types, and the number of twists between layers, a large family of systems falls under the category of moir\'e materials~\cite{Mono-Bi-graphene1,Park2021-tTLG1,Liu2020-tDBLG1,tDBLG2,He2021-tDBLG3,KB_transport_tBG,Adak2022-tDBLG6,Chakraborty_2022,Sinha2022,KB_tDBLG,TMD-homobilayers1,KB_NH_hBN_tBG,rTTLG+hBN1}. In this work, we mainly focus on tBLG and twisted double bilayer graphene (\ie tDBLG-formed when two bilayer graphene systems are rotated with respect to each other by an angle). Concerning band topology, while both tBLG and tDBLG exhibit valley Hall insulating phases, the topological phase diagram of tDBLG is often richer than that of tBLG, as tDBLG can host a broader range of Chern numbers (e.g., $C = \pm 3, \pm 2, \pm 1$~\cite{Koshino-tDBLG,Mohan2021,KB_tDBLG}) compared to the typically lower Chern numbers (such as $C = \pm 1$~\cite{senthil_hBN, Linder_floq, KB_NH_hBN_tBG})) realized in tBLG under similar symmetry-breaking mechanisms.

%SC in graphene, tBLG & tDBLG theory and experiment
%General review on Chiral topological superconductivity(TSC) and proposed TSCs in Graphene
Theoretically, there exists several proposals for the realization of topological superconducting phases in condensed matter systems, among which the most significant three are the following: $(i)$ consideration of a conventional $s$-wave superconductor in close proximity to the two-dimensional (2D) surface state of a three-dimensional (3D) topological insulator and ferromagnetic insulator~\cite{FuKane_2008}, $(ii)$ an $s$-wave superconductor in proximity to a semiconductor with Rashba spin-orbit coupling (RSOC) and a time-reversal-symmetry-breaking mass term~\cite{Sau_2010,Sato_2010}, and $(iii)$ an $s$-wave superconductor in proximity to a quantum Hall or quantum anomalous Hall insulator~\cite{TSC_QH_Zhang}. Following the last proposal, a chiral topological superconducting phase has also been predicted in graphene~\cite{TSC_graphene1,TSC_graphene2}. On the other hand, the nature of intrinsic superconducting states in various graphene systems has been extensively studied using different theoretical methods (\eg mean-field theory, Monte Carlo simulations, and renormalization group analysis) etc., predicting the existence of unconventional $p$-, $d$-, and $f$-wave states~\cite{MFT_Black-Schaffer,MFT_Roy,RG_Nandkishore,RG_Wolf,QMC_Ma,QMC_Dai}, in addition to the conventional $s$-wave state. Earlier studies have investigated the topological properties using the most energetically favorable and symmetry-allowed unconventional pairings in monolayer, bilayer, and trilayer graphene~\cite{TSC_graphene_Pangburn1,TSC_graphene_Pangburn2}. In twisted systems, numerous theoretical proposals aim to characterize the nature of unconventional superconducting pairing using various techniques such as mean-field theory, renormalization group methods, and quantum Monte Carlo simulations~\cite{DMFT_TSC_tBLG,RG_TSC_tBLG,RG_TSC_tDBLG,MFT_TSC_tBLG,QMC_TSC_tBLG,BdG_TSC_tBLG,Bitan_unconv_SC_tBLG}, predicting $p$-, $d$-, and $g$-wave states. In other studies, the topological properties of tBLG with chiral superconductivity have been explored~\cite{topo_TSC_tBLG,HO_topo_TSC_tBLG,TSC_Khosravian_2024} using various different models. However, in the case of tDBLG, no such study-\ie exploring the topological properties of tDBLG in the presence of unconventional pairing
has been reported in the literature to the best of our knowledge.

%---------------Our findings-------------%
In this work, we aim to understand the topological properties of tBLG and tDBLG, considering the chiral $p_{x}+i p_{y}$ superconducting pairing in each layer of graphene. To construct the low-energy model Hamiltonian, we consider the symmetry-allowed order parameter for the $p$-wave pairing in graphene and perform a Taylor series expansion of the pairing function near the valley-$K$ and $K^{'}$ points. Upon writing the Bogoliubov–de Gennes (BdG) Hamiltonian, we calculate the band dispersion for both tBLG and tDBLG. We then discuss how different parameters, such as the superconducting order parameter and chemical potential, affect the system band properties. In the latter part, we focus on the topological properties of both the systems. We calculate the appropriate topological invariant Chern number in different regions of the parameter space. To establish the observed topological phase transitions, we compute the direct band gap and examine the gap-closing points on the moir\'e Brillouin zone (mBZ). 

To understand how varying the twist angles impact the topological superconducting properties, we compute the Chern number in the plane of chemical potential–twist angle for a fixed value of the superconducting order parameter. A rich phase diagram with multiple topological phases is obtained for both tBLG and tDBLGs. Furthermore, to explore the topological properties near the magic angle in tBLG, we analyze the chemical potential–superconducting order parameter plane at the magic angle. A similar phase diagram is also constructed for both the AB-AB and AB-BA stackings of tDBLGs at a twist angle $\theta = 1.3^{\circ}$. In all three systems (tBLG, AB-AB, and AB-BA stacked tDBLGs), we also calculate the squared amplitude of the Bloch states in different topological regions and observe their evolution. Finally, we include the effect of trigonal warping in both AB-AB and AB-BA tDBLGs and calculate the Chern number in the chemical potential–superconducting order parameter plane at $\theta = 1.3^{\circ}$. The resulting phase diagram reveals new phases with higher Chern numbers over an extended parameter range and exhibits a reduction in the topologically trivial region compared to the previous results.

%-------------- Structure --------------%
The remainder of the article is organized as follows. In Sec.~\ref{Sec:II}, we introduce the model Hamiltonian for tBLG and tDBLG in the presence of a chiral $p_{x} + i p_{y}$ superconducting pairing. In Sec.~\ref{Sec:III}, we present the band structure corresponding to the BdG Hamiltonian for tBLG and tDBLG for different values of the superconducting order parameter and chemical potential. Sections~\ref{Sec:IV} and~\ref{Sec:V} are devoted to the discussion of the topological superconducting properties of tBLG and tDBLG, respectively. In these sections, we discuss various topological phase diagrams to understand the effects due to the variation of twist angle, chemical potential, and superconducting order parameter. To establish the observed superconducting topological phases, we discuss the direct band gap, along with the band structure and squared amplitude of the Bloch states. In Sec.~\ref{Sec:VI}, the effect of trigonal warping on the topological properties is discussed. In Sec.~\ref{Sec:VII}, by introducing a unitary transformation, we demonstrate that unconventional $p+ip$ superconductivity can be realized from a conventional $s$-wave superconductor in presence of a magnetic field and RSOC. Finally, we summarize and conclude our work in Sec.~\ref{Sec:VIII}.

%======================================================
\section{Model and Method}\label{Sec:II}
%======================================================
In this section, we describe the low energy continuum model that we use for investigating the chiral topological superconductivity in both tBLG and tDBLG. In the latter part, we discuss the methods that we employ to characterize the topological properties of such a system. 
%-----------------------------------------------------
\subsection{Model Hamiltonian}\label{subsec:IIA}
%------------------------------------------------------

%-------------------------------------------------
\subsubsection{\rm{tBLG}}
Small-angle twisted systems are best described by an effective low-energy continuum model~\cite{MacDonald-tBLG,Koshino-tBLG}. Here, we write the model Hamiltonian for tBLG in the basis $\Phi_{K_{\xi},\bf{q}} = (A_{q_1}, B_{q_1}, A_{q_2}, B_{q_2})^{T}$ as,
\begin{eqnarray}
	H^{0,\xi}_{\text{tBLG}}({ \mathbf q})=\left(\begin{array}{cc}%
		h^{\xi}_{0}({\mathbf{q_1}}) & T^{\xi}\\
		{T^{\xi}}^\dagger & h^{\xi}_{0}({\mathbf{q_2}}) 
	\end{array}\right)\ ,
	\label{Eq:tBLG}
\end{eqnarray}
with $A_{q_l}$ and $B_{q_l}$ denote the annihilation operators with momentum $q_l$ in layer-$l$ at sublattice-$A$ and $B$, respectively. Also, $h^{\xi}_{0}({\mathbf{q_1}})$ and $h^{\xi}_{0}({\mathbf{q_2}})$ are Hamiltonians for two monolayers of tBLG with $\mathbf{q}_{(l)} = R[(-1)^{(l)} (\theta/2)]\mathbf{q}$ near valley-$K_\xi$. Here, $T^{\xi}$ describes the interlayer coupling between the two layers. We note that under time-reversal symmetry $\mathcal{T}$, the Hamiltonian satisfies $[ H^{0,\xi}_{\text{tBLG}}(\mathbf{q})]^* = H^{0,-\xi}_{\text{tBLG}}(-\mathbf{q})$. This implies that the valley-$K$ ($\xi = +1$) can be mapped to its time-reversal partner, the valley-$K'$ ($\xi = -1$). See Appendix~\ref{AppD} for details of the continuum model and the corresponding parameter values used.

%Superconducting part:	

Considering the tight-binding model of monolayer graphene one can calculate the form factors (see Appendix. \ref{AppA} for derivation) for the nearest neighbor (nn) (see Fig.\ref{mBZ2} (a)) $p_{x}$- and $p_{y}$- superconducting pairing as~\cite{TSC_graphene_Pangburn1, TSC_graphene_Pangburn2},
%\begin{eqnarray}
\begin{align}
	f^{x}_{nn}({\bf k}) &= i \sqrt{2} \Delta_{\mathrm{sc}} e^{\frac{1}{2} i k_{y}} \sin\left(\frac{\sqrt{3}}{2}
	k_{x}\right)\ , \\
	f^{y}_{nn}({\bf k}) &= \frac{2\Delta_{\mathrm{sc}}}{\sqrt{6}} e^{- i k_{y}} \left[ 1 - e^{\frac{3i}{2} k_{y}}\text{cos}\left(\frac{\sqrt{3}}{2} k_{x}\right) \right]\ ,
\end{align}
However, in the context of the low-energy continuum model, we deduce the corresponding form factors near valley-$\mathbf{K_{\xi}}$ (by $\mathbf{k} = \mathbf{K_{\xi}} + \mathbf{q}$). Employing these low-energy form factors, we construct the BdG Hamiltonian for both tBLG and tDBLG systems. A detailed derivation of the low energy version of BdG Hamiltonian for monolayer graphene is also discussed in Appendix~\ref{AppB}.

%--------BdG Hamiltonian----------%
Then, we consider writing the total Hamiltonian in terms of the BdG-Hamiltonian 
considering the following Nambu basis, $\Psi_{\bf{q}} = \bigl( \Phi_{K,\bf{q}\uparrow}, \Phi_{K,\bf{q}\downarrow},\Phi^{\dagger}_{K^{\prime},-\bf{q}\uparrow}, \Phi^{\dagger}_{K^{\prime},-\bf{q}\downarrow} \bigr)^{T}$, such that the total Hamiltonian reads
\begin{align}
	H = \sum_{\mathbf{q}}\Psi_{\bf{q}}^{\dagger} H_{\text{BdG}}(\mathbf{q}) \Psi_{\bf{q}}\ ,
	\label{Eq:tBLG:BdG1}
\end{align}
and the BdG Hamiltonian in the above basis can be written as,
\begin{align}
	H_{\text{BdG}}(\mathbf{q}) = 
	\left( \begin{array}{cc}
		H_{\text{tBLG}}(\mathbf{q}) & H_{\Delta}(\mathbf{q})\\
		H_{\Delta}(\mathbf{q})^\dagger & -H_{\text{tBLG}}(\mathbf{q})) 
	\end{array}\right)\ ,
	\label{Eq:tBLG:BdG2}
\end{align}
where, $H_{\text{tBLG}}(\mathbf{q})$ represents the model Hamiltonian for tBLG at valley-$K$, while $-H_{\text{tBLG}}(\mathbf{q})$ corresponds to the tBLG at valley-$K^{\prime}$ after performing the particle-hole transformation (\ie the hole block is the same as particle block with a minus sign). The off-diagonal matrix $H_{\Delta}(\mathbf{q})$ denotes the superconducting pair potential.

%As, under time-reversal symmetry $(H^{K^{\prime}}_{\text{tBLG}}(\mathbf{-q}))^{*} = H^{K}_{\text{tBLG}}(\mathbf{q})$

The explicit expression for them are given by,
\begin{align}
	H_{\text{tBLG}}(\mathbf{q}) = \sigma_{0} \otimes H^{0}_{\text{tBLG}}(\mathbf{q})\ ,\\
	H_{\Delta}(\mathbf{q}) =\left(\begin{array}{cc}%
		H^{\uparrow}_{\Delta}(\mathbf{q}) & 0\\
		0 & H^{\downarrow}_{\Delta}(\mathbf{q})
	\end{array}\right)\ .
	\label{Eq:tBLG:BdG3}
\end{align}	
%to avoid notational complexity we intentionally omit the valley index $\xi$ 
with $\sigma$ represents the Pauli matrix for spin and,
\begin{align}
	H^{s_z}_{\Delta}(\mathbf{q}) = \left(\begin{array}{cc}%
		h^{s_z}_{\Delta}({\bf q_1}) & \mathbf{0}_{2 \times 2}\\
		\mathbf{0}_{2 \times 2} & h^{s_z}_{\Delta}({\bf q_2}) 
	\end{array}\right)\ .
	\label{Eq:tBLG:BdG4}
\end{align}	

In Eq.~(\ref{Eq:tBLG:BdG4}), $H^{s_z}_{\Delta}$ is written for a particular spin $s_z \in \{\uparrow,\downarrow\}$ and has dimension same as $H^{0}_{\text{tBLG}}$. From Appendix~\ref{AppB} we have the low energy version of the superconducting pairing in 
each layer as following,
\begin{align}
	h^{\uparrow}_{\Delta}({\bf q})= 
	\left( \begin{array}{cc}
	0 & \frac{1}{2} i \sqrt{6} \Delta_{\mathrm{sc}}\\
	-\frac{i \sqrt{3}}{2\sqrt{2}} \Delta_{\mathrm{sc}} q_{+} & 0
	\end{array}\right)\ ,\\
	h^{\downarrow}_{\Delta}({\bf q})= 
	\left( \begin{array}{cc}
		0 & \frac{i \sqrt{3}}{2\sqrt{2}} \Delta_{\mathrm{sc}} q_{+}\\
		-\frac{1}{2} i \sqrt{6} \Delta_{\mathrm{sc}} & 0
	\end{array}\right)\ ,
	\label{Eq:tBLG:BdG5}
\end{align}	
where, $\Delta_{\mathrm{sc}}$ represents the superconducting order parameter and the pairings exist only between the sub-lattices. Also note that, from Eq.~(\ref{Eq:tBLG:BdG3}) and Eq.~(\ref{Eq:tBLG:BdG4}) there is neither spin flip term, nor inter-layer coupling term in the superconducting part of the Hamiltonian.

As in the above BdG Hamiltonian for $p + ip$ superconducting tBLG we do not have any spin-flip term, all the bands are spin degenerate. Hence, in our numerical calculation we consider only one spin sector and discuss our findings. The corresponding BdG-Hamiltonian in the basis $\bigl( \Phi_{K,\bf{q}\hspace{1pt} s_z}, \Phi^{\dagger}_{K^{\prime},-\bf{q} \hspace{1pt} s_z} \bigr)^{T}$ can be written as, 
\begin{align}
	H^{s_z}_{\text{BdG}}(\mathbf{q}) = 
	\left( \begin{array}{cc}
		H^{0}_{\text{tBLG}}(\mathbf{q}) & H^{s_z}_{\Delta}(\mathbf{q})\\
		\big( H^{s_z}_{\Delta}(\mathbf{q}) \big)^\dagger & -H^{0}_{\text{tBLG}}(\mathbf{q})
	\end{array}\right)\ .
	\label{Eq:tBLG:BdG6}
\end{align}

%-------------------------------------------------
\subsubsection{\rm{tDBLG}}
\begin{figure}[h]
	\centering
	\subfigure{\includegraphics[width=0.49\textwidth]{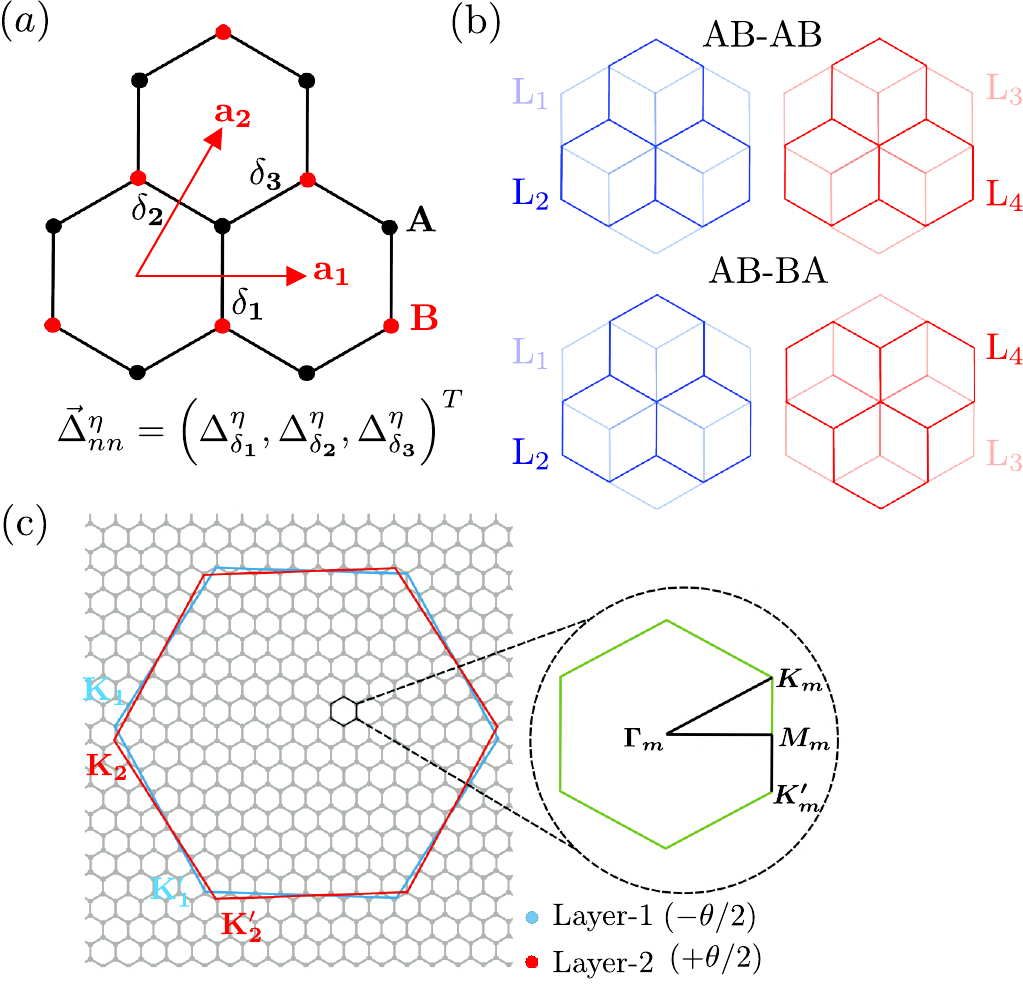}}
	\caption{(a) Schematic of a single layer graphene where $A$, $B$-sublattices are denoted by black and red color dots respectively. Here, $\eta = x,y,z$ denotes different triplet pairing channels, $\delta_{1}$, $\delta_{2}$,
	$\delta_{3}$ denote the nearest neighbor vectors and $\mathbf{a}_{1}$ and $\mathbf{a}_{2}$ represent the lattice vectors. (b) Schematic representation of the double bilayer graphene with AB-AB and AB-BA stackings. Here, $L_{1}$, $L_{2}$, $L_{3}$, $L_{4}$ respectively represent the layer-1, 2 belonging to the first bilayer (shown in blue color) and layer-3, 4 belonging to the second bilayer (shown in red color). (c) The red and cyan hexagons denote the Brillouin zone of the two rotated layers centering each other by $+\theta$/2 and $-\theta$/2. The smaller hexagonal tiling is the folded Brillouin zone \ie the mBZ. One of the smaller hexagons (the moir\'e Brillouin zone) is zoomed in for clarity and shown with the high symmetry path along which the band structure is calculated.
}
	\label{mBZ2}
\end{figure}

%-------------------------------------------------
When two bilayer graphene sheets are stacked on top of each other and slightly rotated, 
it forms a tDBLG. However, depending on the type of stacking of the bilayers, the double bilayer exhibits two stable configurations: one is AB-AB and the other one is AB-BA (see Fig.~\ref{mBZ2}(b)). Below we write the low energy continuum model for both types of tDBLG near valley-$K_{\xi}$ in the basis $\chi_{K_{\xi},\bf{q}} = (A^{1}_{q_1}, B^{1}_{q_1}, A^{2}_{q_1}, B^{2}_{q_1}, A^{3}_{q_2}, B^{3}_{q_2}, A^{4}_{q_2}, B^{4}_{q_2})^{T}$ as,
\begin{eqnarray}
	H^{0,\xi}_{\text{AB-AB}}=\left(\begin{array}{cc}%
		H^{0,\xi}_{AB}({\mathbf{q_1}}) & \tilde{T}^{\xi}\\
		(\tilde{T}^{\xi})^\dagger & H^{0,\xi}_{AB}({\mathbf{q_2}}) 
	\end{array}\right)\ ,
	\label{Eq:ABAB_tDBLG}
\end{eqnarray}

\begin{eqnarray}
	H^{0,\xi}_{\text{AB-BA}}=\left(\begin{array}{cc}%
		H^{0,\xi}_{AB}({\mathbf{q_1}}) & \tilde{T}^{\xi}\\
		(\tilde{T}^{\xi})^\dagger & H^{0,\xi}_{BA}({\mathbf{q_2}}) 
	\end{array}\right)\ ,
	\label{Eq:ABBA_tDBLG}
\end{eqnarray}
where, $A^{j}_{q_l}$ and $B^{j}_{q_l}$ correspond to the annihilation operators with momentum $q_l$ in layer-$j$ and at sublattice-$A$ and $B$, respectively. Here, $q_1$ 
and $q_2$ belong to the rotated BZ of first and second bilayer graphene respectively. Further, $H^{0,\xi}_{AB}({\mathbf{q}})$ and $H^{0,\xi}_{BA}({\mathbf{q}})$ represent the Hamiltonians for two bilayers of tDBLG with AB and BA stacking respectively. Here, we assume that the first bilayer is rotated by $-\theta/2$ and the second bilayer follows a rotation of $\theta/2$ with respect to each other.  
To discuss our results in the latter text, we consider both the situation \ie without (minimal model of tDBLG) and with trigonal warping. $\tilde{T}^{\xi}$ describes the effect of twist in between the layer-2 and layer-3 of the four layers of tDBLG. Here also note that, under time-reversal symmetry $\mathcal{T}$, the Hamiltonian at valley-$K$ ($\xi = +1$) is mapped to its time-reversal partner, the valley-$K'$ ($\xi = -1$). See Appendix~\ref{AppD} for details of the continuum model and the corresponding parameter values used.
%---------------------------------------------------------------------------------------------
%---------------------------------------------------------------------------------------------
%---------------------------------------------------------------------------------------------
%---------------------------------------------------------------------------------------------

%However, to obtain the low energy bands, one needs to go beyond this approximation. For that reason, one can also think that the above Hamiltonian [Eq.~(\ref{Eq:H:AB-AB-Full})] is written in the first mBZ. In our analysis, we consider upto the seventh nearest neighbor and construct the Hamiltonian in the corresponding basis to diagonalize it numerically. 

%Superconducting part:
Now, we write the total Hamiltonian in terms of the BdG Hamiltonian in the following basis, $\Psi_{\bf{q}} = \bigl( \chi_{K,\bf{q} \uparrow}, \chi_{K,\bf{q}\downarrow}, \chi^{\dagger}_{K^{\prime},-\bf{q} \uparrow},\chi^{\dagger}_{K^{\prime},-\bf{q} \downarrow} \bigr)^{T}$ such that, %such that the total Hamiltonian, 
\begin{align}
	H = \sum_{\mathbf{q}}\Psi_{\bf{q}}^{\dagger} H_{\text{BdG}}(\mathbf{q}) \Psi_{\bf{q}}\ ,
	\label{Eq:tDBLG:BdG1}
\end{align}	
and the corresponding BdG Hamiltonian in the above basis can be written as,
\begin{align}
	H_{\text{BdG}}(\bf{q}) = 
	\left( \begin{array}{cc}
		H_{\text{tDBLG}}(\bf{q}) & \tilde{H}_{\Delta}(\bf{q})\\
		\tilde{H}_{\Delta}^\dagger(\bf{q}) & -H_{\text{tDBLG}}(\bf{q}) 
	\end{array}\right)\ ,
	\label{Eq:tDBLG:BdG2}
\end{align}
where, $H_{\text{tDBLG}}$ and $\tilde{H}_{\Delta}$ are the model Hamiltonians for tDBLG and superconducting pairing potential in the low energy approximation and are given by,
\begin{align}
	H_{\text{tDBLG}}(\bf{q}) &= \sigma_{0} \otimes H^{0}_{\text{AB-AB/AB-BA}}(\bf{q})\ ,\\
	\tilde{H}_{\Delta}(\mathbf{q}) &=\left(\begin{array}{cc}%
		\tilde{H}^{\uparrow}_{\Delta}(\bf{q}) & 0\\
		0 & \tilde{H}^{\downarrow}_{\Delta}(\bf{q})
	\end{array}\right)\ .
	\label{Eq:tDBLG:BdG3}
\end{align}	
with $\sigma$ refers the Pauli matrix in spin space and,
\begin{align}
	\tilde{H}^{s_{z}}_{\Delta}(\mathbf{q}) = \left(\begin{array}{cc}%
		\tilde{h}^{s_{z}}_{\Delta}({\bf q_1}) & \mathbf{0}_{2 \times 2}\\
		\mathbf{0}_{2 \times 2} & \tilde{h}^{s_{z}}_{\Delta}({\bf q_2}) 
	\end{array}\right)\ ,
	\label{Eq:tDBLG:BdG4}
\end{align}	
where, the $\tilde{H}^{s_{z}}_{\Delta}$ is written in the moire basis for a spin $s_{z} \in \{\uparrow,\downarrow\}$ and carry dimension same as $H^{0}_{\text{tDBLG}}$. Also, $\tilde{h}^{s_{z}}_{\Delta}({\bf q}) = I_{2 \times 2} \otimes h^{s_{z}}_{\Delta}({\bf q})$ is a 
$4 \times 4$ matrix for the bilayer graphene.

Similar to tBLG, in tDBLG too, we do not have any spin-flip term. Hence, in our numerical calculation we consider only one spin sector. The corresponding BdG-Hamiltonian in the basis $\bigl( \chi_{\bf{q}\hspace{1pt} s_z}, \chi^{\dagger}_{-\bf{q} \hspace{1pt} s_z} \bigr)^{T}$ can be written as, 
\begin{align}
	H^{s_z}_{\text{BdG}}(\mathbf{q}) = 
	\left( \begin{array}{cc}
		H^{0}_{\text{AB-AB/AB-BA}}(\bf{q}) & \tilde{H}^{s_z}_{\Delta}(\mathbf{q})\\
		\big( \tilde{H}^{s_z}_{\Delta}(\mathbf{q}) \big)^\dagger & -H^{0}_{\text{AB-AB/AB-BA}}(\bf{q}) 
	\end{array}\right)\ .
	\label{Eq:tDBLG:BdG5}
\end{align}

In both tBLG and tDBLG, the interlayer coupling, governed by van der Waals interactions, is significantly weaker than the characteristic intralayer energy scale. Motivated by this distinction in energy scales, we restrict our analysis to direct superconducting pairing within individual layers, which is expected to provide the dominant contribution.
	
%we restrict our analysis to superconducting pairing within individual layers, which is expected to provide the leading contribution.}

%-------------------------------------------------------------------------------------------
\subsection{Direct band gap, Bloch states and Chern number}\label{subsec:IIB}
%-------------------------------------------------------------------------------------------
We begin by introducing the key quantities that we use in the subsequent analysis: the direct band gap, Bloch states of the moir\'e system, and the Chern number.

The direct band gap is defined as the minimum energy difference between the first conduction band and the first valence band, considered over the mBZ. Quantitatively, one writes
\begin{equation}
	\delta_{\rm dir} = \min_{\mathbf k\in \rm mBZ} \bigl[\,\epsilon_{1}(\mathbf k) - \epsilon_{-1}(\mathbf k)\bigr]\ ,
	\label{Eq:direct_band_gap}
\end{equation}
where, $\epsilon_{-1}(\mathbf k)$ and $\epsilon_{1}(\mathbf k)$ denote respectively the energy of the highest valence band and the lowest conduction band at wave‐vector $\mathbf k$.

We express the Bloch eigenstates of the moiré Hamiltonian near valley-$K$ and spin-$s_{z}$ as,
\begin{equation}
	\Psi^{\alpha;s_{z}}_{n\mathbf k}(\mathbf r)
	= \sum_{\mathbf G^{m}} C^{\alpha;s_{z}}_{n\mathbf k}(\mathbf G^{m}) \; e^{\,i\,(\mathbf k+ \mathbf G^{m})\!\cdot\mathbf r}\ ,
	\label{Eq:Bloch_state}
\end{equation}
where, $\mathbf k$ is the Bloch wave‐vector, $\mathbf G^{m}$ denotes the reciprocal lattice vectors of the moiré superlattice, $n$ labels the band index, and $\alpha$ labels the sublattice and layer degrees of freedom: for instance in tBLG we have $\alpha\in\{A_{1},B_{1},A_{2},B_{2}\}$, while in tDBLG one has $\alpha\in\{A_{1},B_{1},A_{2},B_{2},A_{3},B_{3},A_{4},B_{4}\}$.

In systems with many bands, and in particular when band crossings or near‐degeneracies occur, it is appropriate to adopt the non‐Abelian formulation of Berry connection and curvature to compute the total Chern number of a set of bands (say from the $\mu^{\rm{th}}$ to the $\nu^{\rm{th}}$ band)~\cite{Fukui-chern_no}. 
The Chern number near valley-$K_{\xi}$ and spin-$s_{z}$ can be written as,
\begin{equation}
	C^{\xi}_{s_z}
	= \frac{1}{2\pi} \int_{\rm mBZ} d^{2}\mathbf k\;\mathrm{Tr}\!\bigl(\mathbf F_{\mathbf{k} s_{z}}\!\cdot\hat{\mathbf z}\bigr)\,,
	\label{Eq:Fukui-chern-1}
\end{equation}
where, the Berry curvature matrix $\mathbf F_{\mathbf{k} s_{z}}^{\,\mu \nu}$ is given by \cite{non-Abelian_Berry_curvature_2019}
\begin{align}
	\mathbf F_{\mathbf{k} s_{z}}^{\,\mu \nu}
	&= \nabla_{\mathbf k}\!\times\! \mathbf A_{\mathbf{k} s_{z}}^{\,\mu \nu}
	+ i\bigl[A_{k_x s_{z}},A_{k_y s_{z}}\bigr]^{\mu \nu}\ , \\
	\mathbf A_{\mathbf{k} s_{z}}^{\,\mu \nu}
	&= i\,\langle \psi^{\mu}_{\mathbf{k} s_{z}}\mid \nabla_{\mathbf k}\mid \psi^{\nu}_{\mathbf{k} s_{z}}\rangle\ .
	\label{Eq:Fukui-chern-3}
\end{align}

Here, $|\psi^{\nu}_{\mathbf{k} s_{z}}\rangle$ denotes the eigenstate of the BdG Hamiltonian with band index $\nu$ and spin-$s_{z}$. The above non‐Abelian formulation is valid when the highest band in the set (say $\nu^{\rm{th}}$ band) is separated by a direct gap from the $(\nu+1)^{\rm{th}}$ band. 

For the numerical implementation, we employ the gauge-invariant method proposed by Fukui \textit{et al.}~\cite{Fukui-chern_no} to compute the Chern number. In all topological phase diagrams presented in this work, the Chern numbers are evaluated by summing over all occupied (valence) bands below the charge neutrality point of the BdG Hamiltonian. Furthermore, we explicitly retain the valley index in the definition of the Chern number to indicate that it is computed for the normal-state Hamiltonian corresponding to valley-$K_{\xi}$, as appearing in the electron block of the BdG Hamiltonian for tBLG/tDBLG.

%======================================================
\section{Band Dispersion}\label{Sec:III}
%======================================================
In this section, we explore how the low-energy electronic band dispersion of tBLG and tDBLG evolves with changes in the superconducting order parameter and chemical potential considering the $p+ip$ superconducting pairing.
%~~~~~~~~~~~~~~~~~~~~~~~~~~~~~~~~~~~~~~~~~~~~~~~~~~~~~~~~~
%~~~~~~~~~~~~~~~~~~~~~~~~~~~~~~~~~~~~~~~~~~~~~~~~~~~~~~~~~
%-------------------------------------------------
\subsection{tBLG}
%-------------------------------------------------
%~~~~~~~~~~~~~~~~~~~~~~~~~~~~~~~~~~~~~~~~~~~~~~~~~~~~~~~~~~~~~~~~~~~~~~~~~~~~
%~~~~~~~~~~~~~~~~~~~~~~~~~~~~~~~~~~~~~~~~~~~~~~~~~~~~~~~~~~~~~~~~~~~~~~~~~~~~
\begin{figure*}[t]
	\centering
	\subfigure{\includegraphics[width=1.02\textwidth]{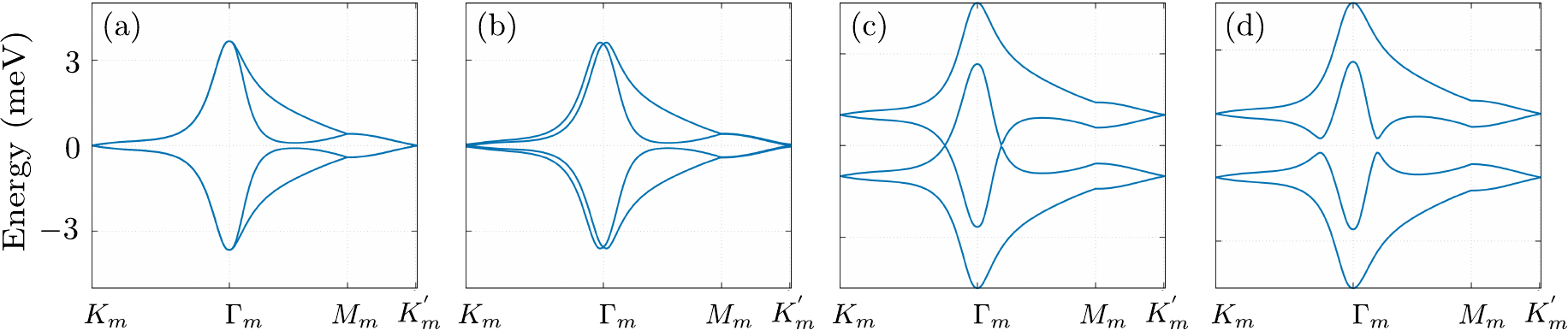}}
	\caption{The electronic band dispersion of tBLG near valley-$K$ along the high symmetry path is demonstrated in presence of different values of chemical potential ($\mu$) and superconducting order parameter ($\Delta_{\mathrm{sc}}$) at twist angle $\theta = 1.05^{o}$. We choose the other model parameters as following, in panel (a) $\Delta_{\mathrm{sc}}$ = 0 meV, $\mu$ = 0 meV, panel (b) $\Delta_{\mathrm{sc}}$ = 5 meV, $\mu$ = 0 meV, panel (c) $\Delta_{\mathrm{sc}}$ = 0 meV, $\mu$ = 1 meV, and 
	panel (d) $\Delta_{\mathrm{sc}}$ = 5 meV, $\mu$ = 1 meV. 
	}
	\label{tBLG_bands}
\end{figure*}
%~~~~~~~~~~~~~~~~~~~~~~~~~~~~~~~~~~~~~~~~~~~~~~~~~~~~~~~~~~~~~~~~~~~~~~~~~~~~~~
%~~~~~~~~~~~~~~~~~~~~~~~~~~~~~~~~~~~~~~~~~~~~~~~~~~~~~~~~~~~~~~~~~~~~~~~~~~~~~~
In Fig.~\ref{tBLG_bands}, we present the electronic band dispersion of tBLG with a $p + ip$ superconducting state near valley-$K$ at a twist angle of $\theta = 1.05^{\circ}$ (magic angle for tBLG~\cite{MacDonald-tBLG,Cao2018-unconv_sc,Origin_of_Magic_Angle-tBLG}) along the high-symmetry path $K_{m} - \Gamma_{m} - M_{m} - K_{m}^{'}$ in the mBZ (as shown in Fig.~\ref{mBZ2}(c)). We analyze the band structure (by diagonalizing the Hamiltonian in Eq.~(\ref{Eq:tBLG:BdG6})) for various values of the chemical potential ($\mu$) and superconducting order parameter ($\Delta_{\mathrm{sc}}$) to gain insight into the system's behavior. 

In the absence of both superconducting pairing ($\Delta_{\mathrm{sc}} = 0$ meV) and at chemical potential $\mu = 0$ meV, as shown in Fig.~\ref{tBLG_bands}(a), 
the moiré flat bands correspond to the bare tBLG Hamiltonian bands arising around both valley-$K$ and valley-$K^{\prime}$. This appears because, in the BdG Hamiltonian, 
the particle and hole sectors correspond respectively to the bare Hamiltonians near valley-$K$ and valley-$K^{\prime}$, as they are related by time-reversal symmetry. In Fig.~\ref{tBLG_bands}(b), we show the effect of turning on the superconducting order parameter ($\Delta_{\mathrm{sc}} = 5$ meV) while keeping $\mu = 0$ meV. We observe that although finite $\Delta_{sc}$ lifts the degeneracies along $K_{m}-\Gamma_{m}$ direction, the flat bands remain gapless under this condition. In contrast, Fig.~\ref{tBLG_bands}(c) illustrates that with a finite chemical potential ($\mu = 1$ meV) and no superconductivity ($\Delta_{\mathrm{sc}} = 0$ meV), a band gap of approximately $|2\mu|$ opens at the $K_{m}$ and $K_{m}^{'}$ points (\ie lifts the degeneracy). Moreover, as the chemical potential increases, the two sets of flat bands separate and move apart. Finally, when both the superconducting order parameter and chemical potential are present, as depicted in Fig.~\ref{tBLG_bands}(d), a band gap emerges between the flat bands, reflecting the combined effect of superconductivity and doping. Further, in Appendix~\ref{AppC}, we present the electronic and particle-hole resolved spectral functions. The particle-hole resolved spectral function exhibits band inversion, signifying the emergence of topological superconducting phase in the system.

%-----------------------------------------------------------------------------------
%-----------------------------------------------------------------------------------

%--------------------------------------------
\subsection{tDBLG}
%--------------------------------------------

%~~~~~~~~~~~~~~~~~~~~~~~~~~~~~~~~~~~~~~~~~~~~~~~~~~~~~~~~~~~~~~~~~~~~~~~~~~~~~~~~~~~~
%~~~~~~~~~~~~~~~~~~~~~~~~~~~~~~~~~~~~~~~~~~~~~~~~~~~~~~~~~~~~~~~~~~~~~~~~~~~~~~~~~~~~
\begin{figure*}[t]
	\centering
	\subfigure{\includegraphics[width=1.0\textwidth]{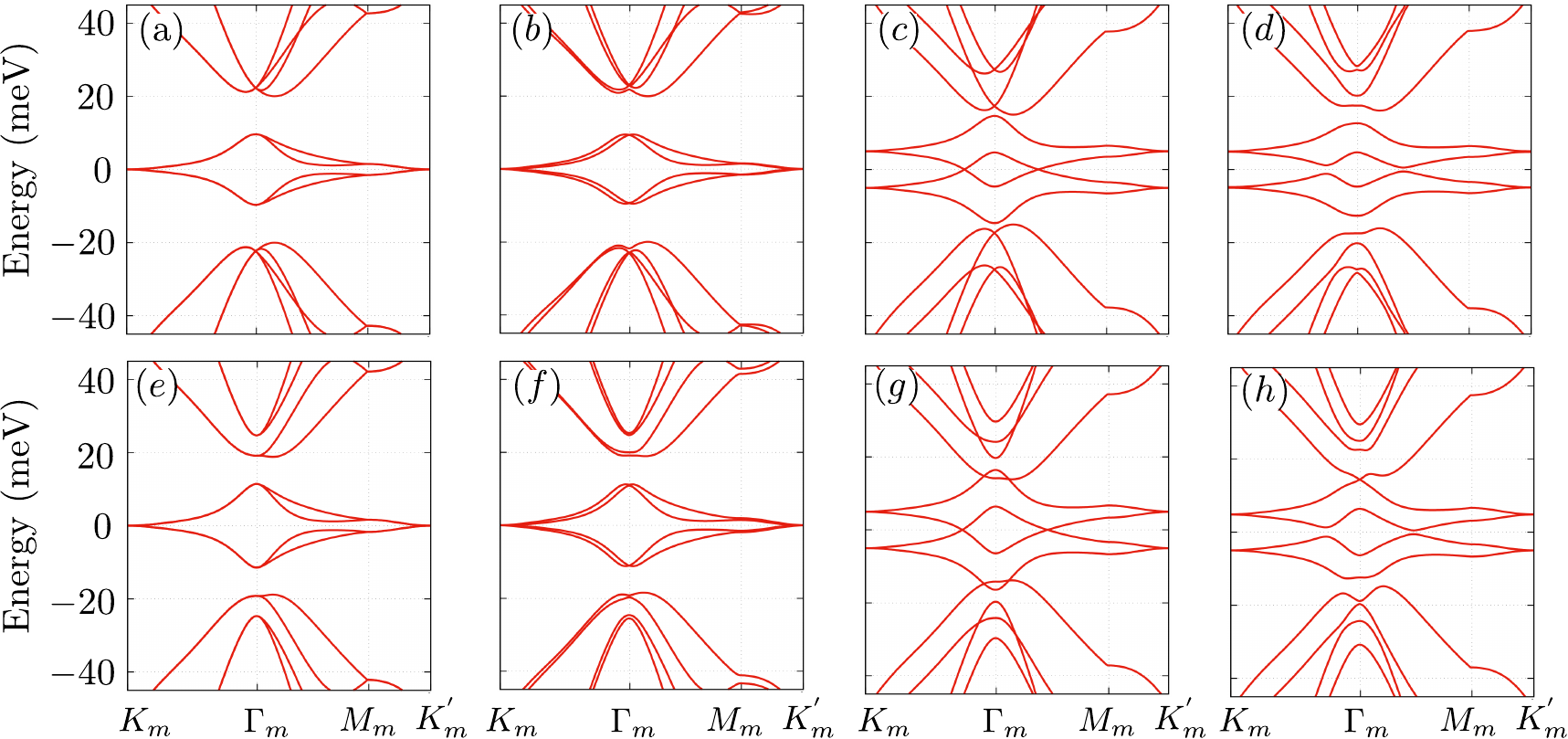}}
	\caption{The electronic band dispersion of tDBLG near valley-$K$ along the high symmetry path is demonstrated in presence of different values of chemical potential ($\mu$) and superconducting pairing ($\Delta_{\mathrm{sc}}$) at the twist angle $\theta = 1.3^{o}$. In the upper panel (\ie (a)-(d)) results are shown for AB-AB tDBLG and in the lower panel (\ie (e)-(h)) for AB-BA tDBLG respectively. We choose the model parameters as following, in panels (a) and (e) $\Delta_{\mathrm{sc}}$ = 0 meV, $\mu$ = 0 meV, 
	panels (b) and (f) $\Delta_{\mathrm{sc}}$ = 10 meV, $\mu$ = 0 meV, panels (c) and (g) $\Delta_{\mathrm{sc}}$ = 0 meV, $\mu$ = 5 meV, and panels (d) and (h) $\Delta_{\mathrm{sc}}$ = 15 meV, $\mu$ = 5 meV. 
	}
	\label{tDBLG_bands}
\end{figure*}
%~~~~~~~~~~~~~~~~~~~~~~~~~~~~~~~~~~~~~~~~~~~~~~~~~~~~~~~~~~~~~~~~~~~~~~~~~~~~~~~~~~~
%~~~~~~~~~~~~~~~~~~~~~~~~~~~~~~~~~~~~~~~~~~~~~~~~~~~~~~~~~~~~~~~~~~~~~~~~~~~~~~~~~~~

Here we discuss the electronic band dispersion of tDBLG  with $p_{x} + ip_{y}$ superconducting pairing near valley-$K$ at a twist angle $\theta = 1.3^{o}$ 
(magic angle for tDBLG~\cite{Haddadi-tDBLG,He2021-tDBLG3}) along the high symmetry path $K_{m} - \Gamma_{m} - M_{m} - K_{m}^{'}$ on the mBZ for both the AB-AB and AB-BA stacking (see Fig.\ref{tDBLG_bands}). Akin to tBLG in tDBLG too, for various values of chemical potential ($\mu$) and superconducting order parameter ($\Delta_{\mathrm{sc}}$) we investigate the band structure to understand the system. In the upper panel (\ie in Figs.~(a)-(d)) band structure is shown for AB-AB tDBLG and in lower panel (\ie in Figs.~(e)-(h)) the same is depicted for AB-BA tDBLG. In absence of both the superconducting order parameter ($\Delta_{\mathrm{sc}}$ = 0 meV) and chemical potential ($\mu$ = 0 meV) we obtain the usual band structure of the flat bands corresponding to the bare Hamiltonians as shown in Fig.~\ref{tDBLG_bands}(a) and Fig.~\ref{tDBLG_bands}(e) respectively for the AB-AB and AB-BA stacked tDBLG. In Fig.~\ref{tDBLG_bands}(b) and Fig.~\ref{tDBLG_bands}(f), we turn on the superconducting order parameter ($\Delta_{\mathrm{sc}}$ = 10 meV) and maintain $\mu$ = 0 meV. Then, we observe that the flat bands remain gapless respectively for AB-AB and AB-BA tDBLG. Note that, like tBLG here too, the degeneracies along $K_{m}-\Gamma_{m}$ is lifted. On the other hand, in absence of the superconducting order parameter ($\Delta_{\mathrm{sc}}$ = 0 meV) and finite value of the chemical potential (\ie $\mu$ = 5 meV) it is observed that at $K_{m}$, $K_{m}^{'}$ point a band gap of the order of $|2\mu|$ appears for both type of the tDBLGs, similar to tBLG. Similarly, it is also seen that with the enhancement of chemical potential
(\ie $\mu$) two sets of flat bands are separated from each other as shown in Fig.~\ref{tDBLG_bands}(c) and Fig.~\ref{tDBLG_bands}(g). Finally, in Fig.~\ref{tDBLG_bands}(d) and Fig.~\ref{tDBLG_bands}(h), when both the superconducting order parameter($\Delta_{\mathrm{sc}} = 10 $ meV) and chemical potential ($\mu = 15$ meV) becomes finite, a band gap between the flat bands is created for both AB-AB and AB-BA tDBLGs respectively. Furthermore, in Appendix~\ref{AppC}, we present the electronic and particle-hole resolved spectral functions for tDBLG. 
The particle-hole resolved spectral function exhibits band inversion, signifying the topological phase transition and emergence of topological superconducting phase therein.

%----------------------------------------------------------------
%----------------------------------------------------------------

%======================================================
\section{Topological superconducting properties of tBLG}\label{Sec:IV}
%======================================================

In this section, we investigate the topological characteristics of tBLG in the presence of chiral \(p+ip\) superconducting pairing.

%~~~~~~~~~~~~~~~~~~~~~~~~~~~~~~~~~~~~~~~~~~~~~~~~~~~~~~~~~
%~~~~~~~~~~~~~~~~~~~~~~~~~~~~~~~~~~~~~~~~~~~~~~~~~~~~~~~~~
%-------------------------------------------------
\subsection{Chern number}
%-------------------------------------------------

%~~~~~~~~~~~~~~~~~~~~~~~~~~~~~~~~~~~~~~~~~~~~~~~~~~~~~~~~~~~~
%~~~~~~~~~~~~~~~~~~~~~~~~~~~~~~~~~~~~~~~~~~~~~~~~~~~~~~~~~~~~
\begin{figure}[h]
	\centering
	\subfigure{\includegraphics[width=0.5\textwidth]{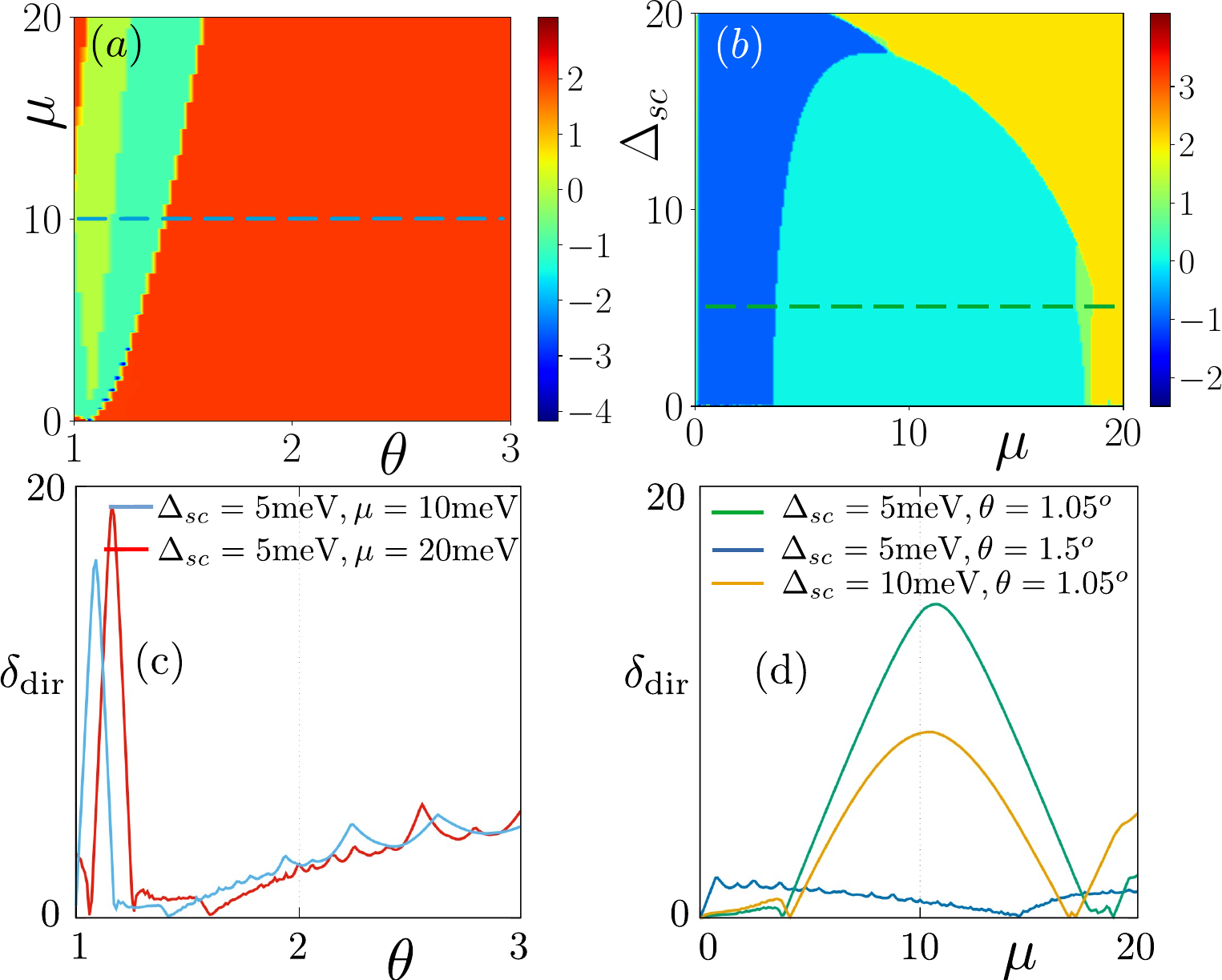}}
	\caption{Density plots and line plots respectively for the topological phase diagram and direct band gap in tBLG with chiral $p_{x} + ip_{y}$ superconductivity around valley-$K$ and spin-up are depicted. In panel (a), Chern number ($\mathcal{C}^{K}_{\uparrow}$) is shown in the plane of twist angle ($\theta$) and chemical potential ($\mu$) for superconducting order parameter ($\Delta_{\mathrm{sc}} = 5$ meV). In panel (b), the same is shown in the chemical potential ($\mu$) and superconducting order parameter ($\Delta_{\mathrm{sc}}$) plane for the twist angle $\theta = 1.05^{o}$. 	
	Direct band gap ($\delta_{\text{dir}}$ in meV) is displayed as a function of twist angle ($\theta$) and chemical potential ($\mu$) respectively in panels 
	(c) and (d) for the same system with various values of twist angle ($\theta$), chemical potential ($\mu$) and superconducting order parameter ($\Delta_{\mathrm{sc}}$), as mentioned in the figures. 	
	}
	\label{tBLG_Chern_phase}
\end{figure}
%~~~~~~~~~~~~~~~~~~~~~~~~~~~~~~~~~~~~~~~~~~~~~~~~~~~~~~~~~~~~~
%~~~~~~~~~~~~~~~~~~~~~~~~~~~~~~~~~~~~~~~~~~~~~~~~~~~~~~~~~~~~~
 
To investigate the topological properties of tBLG in presence of $p_{x} + ip_{y}$ superconducting  pairing, %in twisted bilayer graphene (tBLG), 
we compute the Chern number ($\mathcal{C}^{K}_{\uparrow}$) corresponding to the valley-$K$ sector and spin-up component ($s_{z} = +1$). The system is examined across a suitable range of parameters. 

First, we explore the behavior of Chern number in $(\theta, \mu)$ plane, where $\theta$ is the twist angle and $\mu$ is the chemical potential, considering the superconducting order parameter fixed at $\Delta_{\mathrm{sc}} = 5$~meV, as shown in Fig.~\ref{tBLG_Chern_phase}(a). The twist angle is varied from $\theta = 1^{\circ}$ to $\theta = 3^{\circ}$. Near the magic angle ($\theta = 1.05^{\circ}$), several distinct topological phases emerge: $\mathcal{C}^{K}_{\uparrow} = -1$ phase (cyan region), $\mathcal{C}^{K}_{\uparrow} = +2$ phase (red region), and a topologically trivial phase with $\mathcal{C}^{K}_{\uparrow} = 0$ (green region). Away from the magic angle, for slightly larger twist angles, only the $\mathcal{C}^{K}_{\uparrow} = +2$ phase (red region) remains. This behavior can be attributed to the increasing bandwidth of the flat bands with larger twist angles, while the chemical potential remains comparatively unchanged.

To gain deeper insight, we compute $\mathcal{C}^{K}_{\uparrow}$ in the $(\mu, \Delta_{\mathrm{sc}})$ plane at the magic angle $\theta = 1.05^{\circ}$. This is shown in Fig.~\ref{tBLG_Chern_phase}(b). The resulting phase diagram reveals several distinct topological regions: $\mathcal{C}^{K}_{\uparrow} = -1$ (blue region), $\mathcal{C}^{K}_{\uparrow} = +1$ (green region), and $\mathcal{C}^{K}_{\uparrow} = +2$ (yellow region). A large part of the phase diagram corresponds to the topologically trivial phase, $\mathcal{C}^{K}_{\uparrow} = 0$ (cyan region). Further understanding from the gap closing transitions is presented in 
Figs.~\ref{tBLG_Chern_phase}(c)-(d) (see latter text for discussion).

%%%%%%%%%%%%%%%%%%%%%%%%%%%%%%%%%%%%%%%%%%%%%%%%%%%%%%%%%%%%%%
%~~~~~~~~~~~~~~~~~~~~~~~~~~~~~~~~~~~~~~~~~~~~~~~~~~~~~~~~~
%~~~~~~~~~~~~~~~~~~~~~~~~~~~~~~~~~~~~~~~~~~~~~~~~~~~~~~~~~
\begin{figure*}[]
	\centering
	\subfigure{\includegraphics[width=0.9\textwidth]{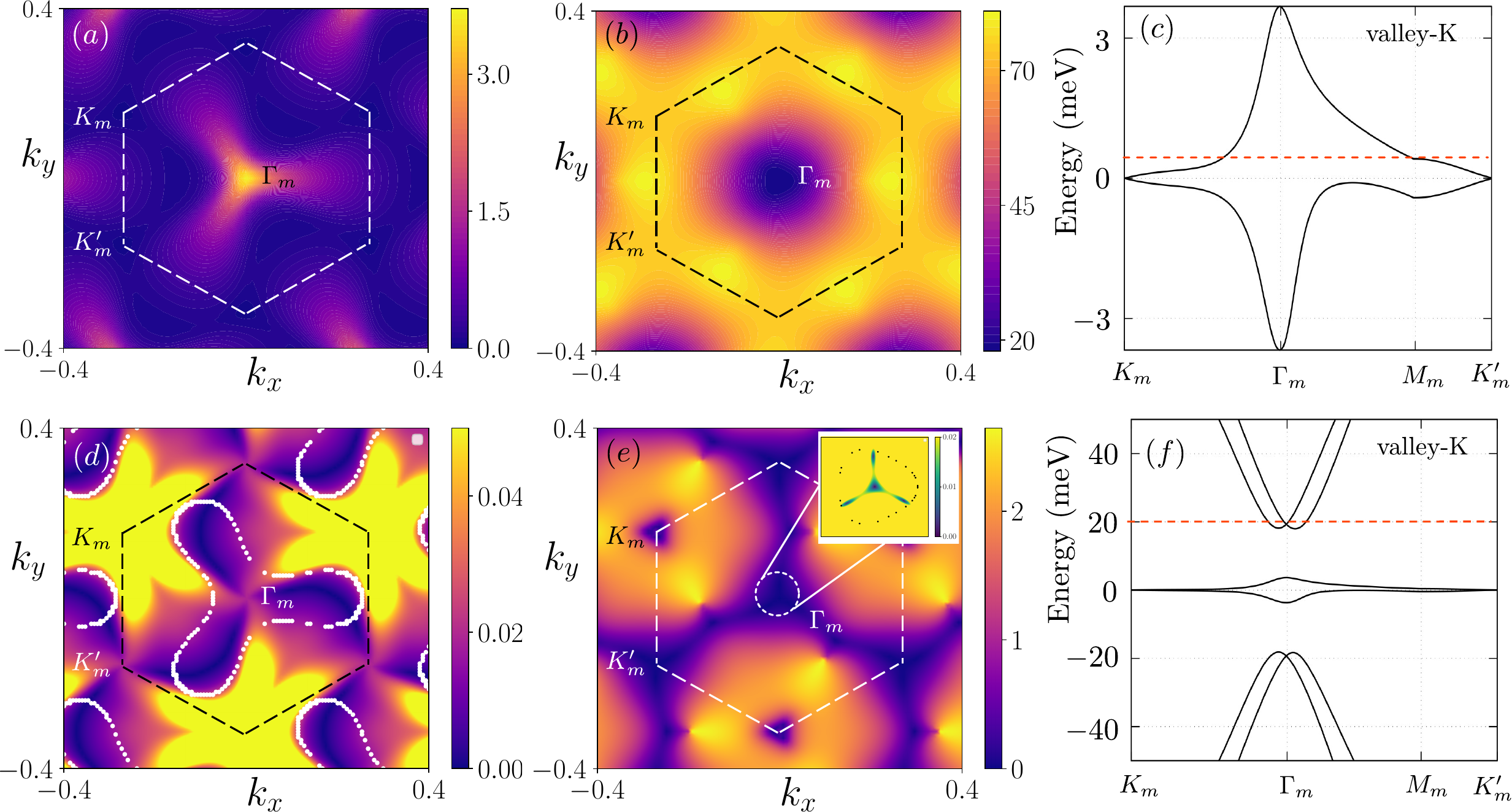}}
	\caption{The normal-state band structure for tBLG at valley-$K$ is shown in Fig.~($a$) and Fig.~($b$), corresponding to the first and second conduction bands in the Brillouin zone, respectively. In Fig.~($c$) and Fig.~($f$), we illustrate the same normal-state band structure along the high-symmetry path for different energy windows to indicate the chemical potentials. Finally, in Fig.~($d$) and Fig.~($e$), we show the projection of the superconducting pairing onto the first and second conduction bands, respectively.
	}
	\label{tBLG_winding}
\end{figure*}
%~~~~~~~~~~~~~~~~~~~~~~~~~~~~~~~~~~~~~~~~~~~~~~~~~~~~~~~~~
%~~~~~~~~~~~~~~~~~~~~~~~~~~~~~~~~~~~~~~~~~~~~~~~~~~~~~~~~~

For chiral topological superconductors in the weak-pairing limit, the Chern number can be determined by the winding of the order parameter around the normal-state Fermi surface (FS)~\cite{TSC_review_Kallin_2016, TSC_graphene_Pangburn1,TSC_graphene_Pangburn2}. Guided by this principle, in Fig.~\ref{tBLG_winding}, we analyze the $\mu$--$\Delta_{\text{sc}}$ phase diagram of tBLG at the magic angle [see Fig.~\ref{tBLG_Chern_phase}(b)], focusing on the regime $\Delta_{\text{sc}} \leq 6$ meV, which lies below the flat-band bandwidth ($\approx 10$ meV). In this limit, the phase boundaries can be directly understood from the evolution of the normal-state FS as a function of the chemical potential ($\mu$). The normal-state band structure at valley-$K$ is shown in Fig.~\ref{tBLG_winding}(c) and Fig.~\ref{tBLG_winding}(f), where the dashed red lines indicate the position of chemical potential at $\mu = 0.5$ and $20$ meV, respectively. In Figs.~\ref{tBLG_winding}(a) and \ref{tBLG_winding}(b), we display the first and second conduction bands in the reciprocal space, with hexagons highlighting the mBZ. To connect the superconducting properties with the normal-state electronic structure, we project the pairing potential $H_{\Delta}(\mathbf{q})$ onto the band basis. For the spin-up sector, the effective pairing is given by
%~~~~~~~~~~~~~~~~~~~~~~~~~~~~~~~~~
\begin{eqnarray}
	\Delta^{\uparrow}_{\text{eff}}(\mathbf{q}) &=& \langle \Phi_{K,\mathbf{q}\uparrow}|H^{\uparrow}_{\Delta}(\mathbf{q})|\Phi^{*}_{K',-\mathbf{q}\uparrow} \rangle \non \\
	&=& \langle \Phi_{K,\mathbf{q}\uparrow}|H^{\uparrow}_{\Delta}(\mathbf{q})|\Phi_{K,\mathbf{q}\uparrow} \rangle\ ,
\end{eqnarray}
%~~~~~~~~~~~~~~~~~~~~~~~~~~~~~~~~~
Here, $\vert \Phi_{K,\mathbf{q}\uparrow} \rangle$ denotes the eigenstate of the normal-state Hamiltonian at valley-$K$ with spin up, while $\vert \Phi^{*}_{K',-\mathbf{q}\uparrow} \rangle$ corresponds to the hole partner. The second line follows from time-reversal symmetry, which relates the $K$ and $K'$ valleys of tBLG. The magnitude of $\Delta^{\uparrow}_{\text{eff}}(\mathbf{q})$ is shown in Fig.~\ref{tBLG_winding}(d) and Fig.~\ref{tBLG_winding}(e), for projections onto the first and second conduction bands, respectively.
%~~~~~~~~~~~
Just above $\mu = 0$, the FS encloses the $\Gamma_{m}$ point, as illustrated by the white dotted contour for $\mu = 0.5$ meV in Fig.~\ref{tBLG_winding}(d). Within the mBZ, three $C_{3}$-symmetric gapless patches are present. The $p+ip$-wave order parameter exhibits a single winding around the FS, yielding a Chern number $C^{K}_{\uparrow} = -1$. As $\mu$ increases, the FS shrinks towards $\Gamma_{m}$. Around $\mu \simeq 3.8$ meV [see Fig.~\ref{tBLG_winding}(c)], it leaves the first conduction band, marking the termination of the $C^{K}_{\uparrow} = -1$ phase. For chemical potentials lying between the two conduction bands, the system becomes topologically trivial with $C^{K}_{\uparrow} = 0$.
%~~~~~~~~~~~
Near $\mu \approx 18$ meV, the chemical potential begins to intersect the second conduction band. At $\mu = 20$ meV, we present $|\Delta^{\uparrow}_{\text{eff}}(\mathbf{q})|$ for this band, together with a zoomed-in view of the gapless region around the $\Gamma_{m}$ point. As shown in the inset, the FS (black dotted line) at $\mu = 20$ meV encloses three symmetry-equivalent gapless points as well as one at $\Gamma_{m}$. Owing to $C_{3}$ symmetry, the three equivalent points collectively contribute one unit to the winding around the Fermi surface, while the central point contributes an additional unit. Consequently, the total Chern number in this regime is $C^{K}_{\uparrow} = 2$.
%~~~~~~~~~~~
We also identify a smaller region in the phase diagram with $C^{K}_{\uparrow} = 1$ (see Fig.~\ref{tBLG_Chern_phase}(b)), which arises when the FS, for $\mu \approx 17$--$20$ meV, encloses only one type of gap-closing point. However, for larger values of $\Delta_{\text{sc}}$, this FS based analysis breaks down. 
This is evident in the region $\Delta_{\text{sc}} \gtrsim 10$ meV, where the phase boundaries no longer follow the evolution of the normal-state FS.

%%%%%%%%%%%%%%%%%%%%%%%%%%%%%%%%%%%%%%%%%%%%%%%%%%%%%%%%%%%%%

%---------------------------------------------------------------------------------

%--------------------------------------------
\subsection{Gap closing transitions}
%--------------------------------------------
%~~~~~~~~~~~~~~~~~~~~~~~~~~~~~~~~~~~~~~~~~~~~~~~~~~~~~~~~~~~~~~~~~~~~~~~~~~~~~~~~~~~~~~~~~
%~~~~~~~~~~~~~~~~~~~~~~~~~~~~~~~~~~~~~~~~~~~~~~~~~~~~~~~~~~~~~~~~~~~~~~~~~~~~~~~~~~~~~~~~~
\begin{figure}[h]
	\centering
	\subfigure{\includegraphics[width=0.5\textwidth]{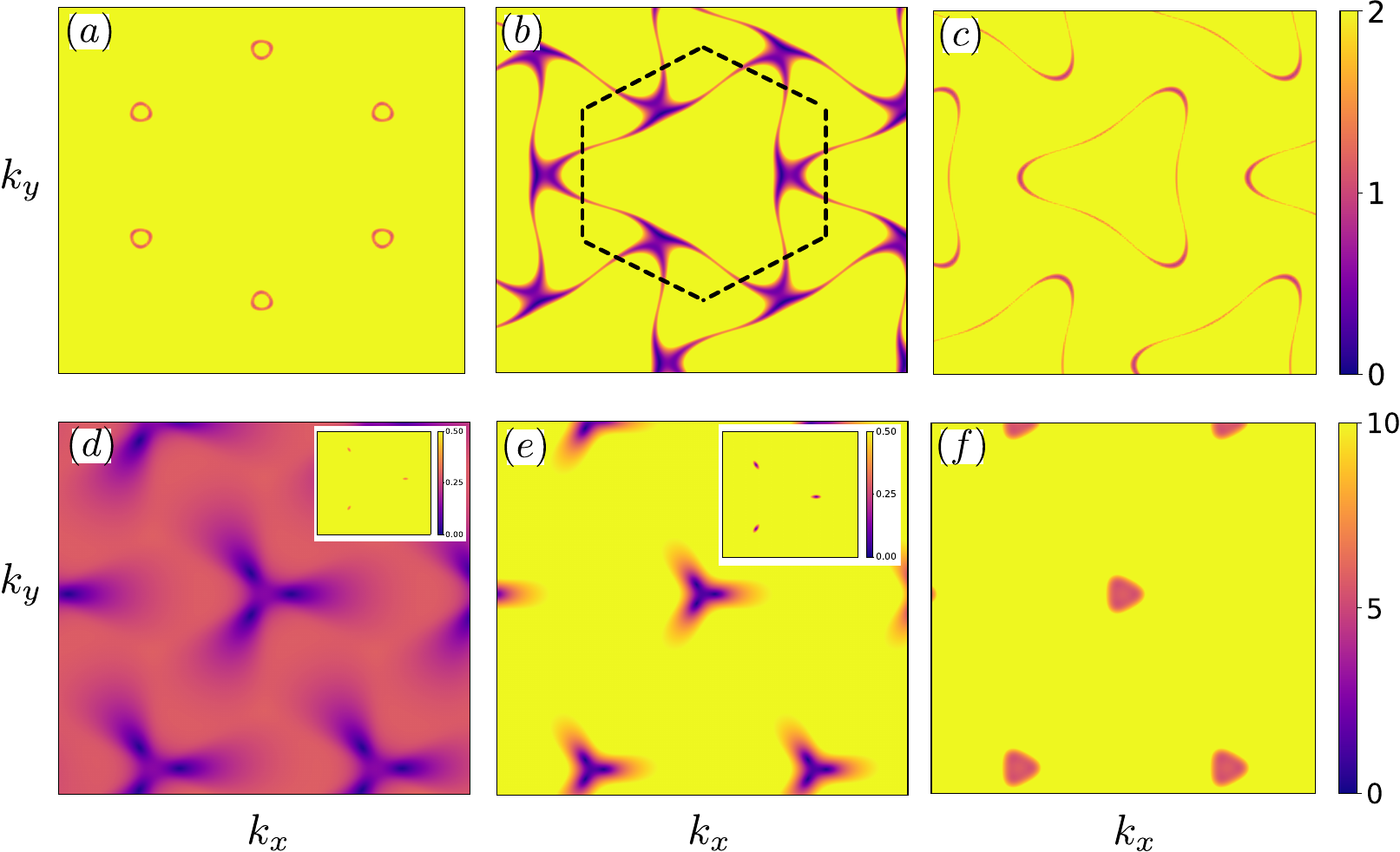}}
	\caption{Density plots for the band gap (in meV) are shown in the $k_{x} - k_{y}$ plane within the mBZ. Panels (a) - (c) correspond to $\mu = 3$ meV, $\mu = 14.5$ meV and $\mu = 20$ meV in case of $p_{x} + ip_{y}$ superconducting tBLG with $\Delta_{\mathrm{sc}} = 5$ meV and twist angle $\theta = 1.5^{o}$ near valley-$K$ and spin-up. On the other hand, panels (d) - (f) correspond to $\mu = 3$ meV, $\mu = 6.3$ meV and $\mu = 15$ meV with $\Delta_{\mathrm{sc}} = 20$ meV and twist angle $\theta = 1.05^{o}$ in the same system.
	}
	\label{tBLG_BZ_gap}
\end{figure}
%~~~~~~~~~~~~~~~~~~~~~~~~~~~~~~~~~~~~~~~~~~~~~~~~~~~~~~~~~~~~~~~~~~~~~~~~~~~~~~~~~~~~~~~~~~
%~~~~~~~~~~~~~~~~~~~~~~~~~~~~~~~~~~~~~~~~~~~~~~~~~~~~~~~~~~~~~~~~~~~~~~~~~~~~~~~~~~~~~~~~~~

To further elucidate the topological phase transitions observed in Figs.~\ref{tBLG_Chern_phase}(a) and (b), we compute the direct band gap, defined in Eq.~(\ref{Eq:direct_band_gap}), for tBLG in presence of chiral $p_{x} + ip_{y}$ superconductivity near the valley-$K$ point and for spin-up states. In Fig.~\ref{tBLG_Chern_phase}(c), we present the variation of the direct band gap ($\delta_{\text{dir}}$) as a function of the twist angle $\theta$, where the two curves correspond to chemical potentials $\mu = 10$~meV and $\mu = 20$~meV, for a fixed superconducting order parameter $\Delta_{\mathrm{sc}} = 5$~meV. These curves correspond to the respective parameter line cuts shown in Fig.~\ref{tBLG_Chern_phase}(a). For $\mu = 10$~meV (cyan line), we identify two distinct topological phase transitions: first from $\mathcal{C}^{K}_{\uparrow} = 0$ to $-1$, and subsequently from $\mathcal{C}^{K}_{\uparrow} = -1$ to $+2$. These transitions are marked by two instances where the direct band gap closes, as can be seen in Fig.~\ref{tBLG_Chern_phase}(c). In contrast, the $\mu = 20$~meV case (red line) exhibits three gap-closing points, corresponding to the transitions $\mathcal{C}^{K}_{\uparrow} = +2 \rightarrow 0$, $\mathcal{C}^{K}_{\uparrow} = 0 \rightarrow -1$, and $\mathcal{C}^{K}_{\uparrow} = -1 \rightarrow +2$. In Fig.~\ref{tBLG_Chern_phase}(d), we illustrate the dependence of $\delta_{\text{dir}}$ on the chemical potential $\mu$. The green line represents the cut at $\Delta_{\mathrm{sc}} = 5$~meV in Fig.~\ref{tBLG_Chern_phase}(b), which captures three successive topological transitions: $\mathcal{C}^{K}_{\uparrow} = -1 \rightarrow 0$, $0 \rightarrow +1$, and $+1 \rightarrow +2$. Similarly, the yellow line corresponds to $\Delta_{\mathrm{sc}} = 10$~meV in the same phase diagram, where the direct gap closes twice, indicating transitions from $\mathcal{C}^{K}_{\uparrow} = -1 \rightarrow 0$ and $0 \rightarrow +2$. Finally, the blue line in Fig.~\ref{tBLG_Chern_phase}(d), corresponding to $\theta = 1.5^{\circ}$ (as in Fig.~\ref{tBLG_Chern_phase}(a)), shows a single gap-closing point, marking a direct topological transition from $\mathcal{C}^{K}_{\uparrow} = +2$ to $-1$.

%~~~~~~~~~~~~~~~~~~~~~~~~~~~~~~~~~~~~~~~~~~~~~~~~~~~~~~~~~~~~~~~~~~~~~
%~~~~~~~~~~~~~~~~~~~~~~~~~~~~~~~~~~~~~~~~~~~~~~~~~~~~~~~~~~~~~~~~~~~~~
\begin{figure*}[t]
	\centering
	\subfigure{\includegraphics[width=0.9\textwidth]{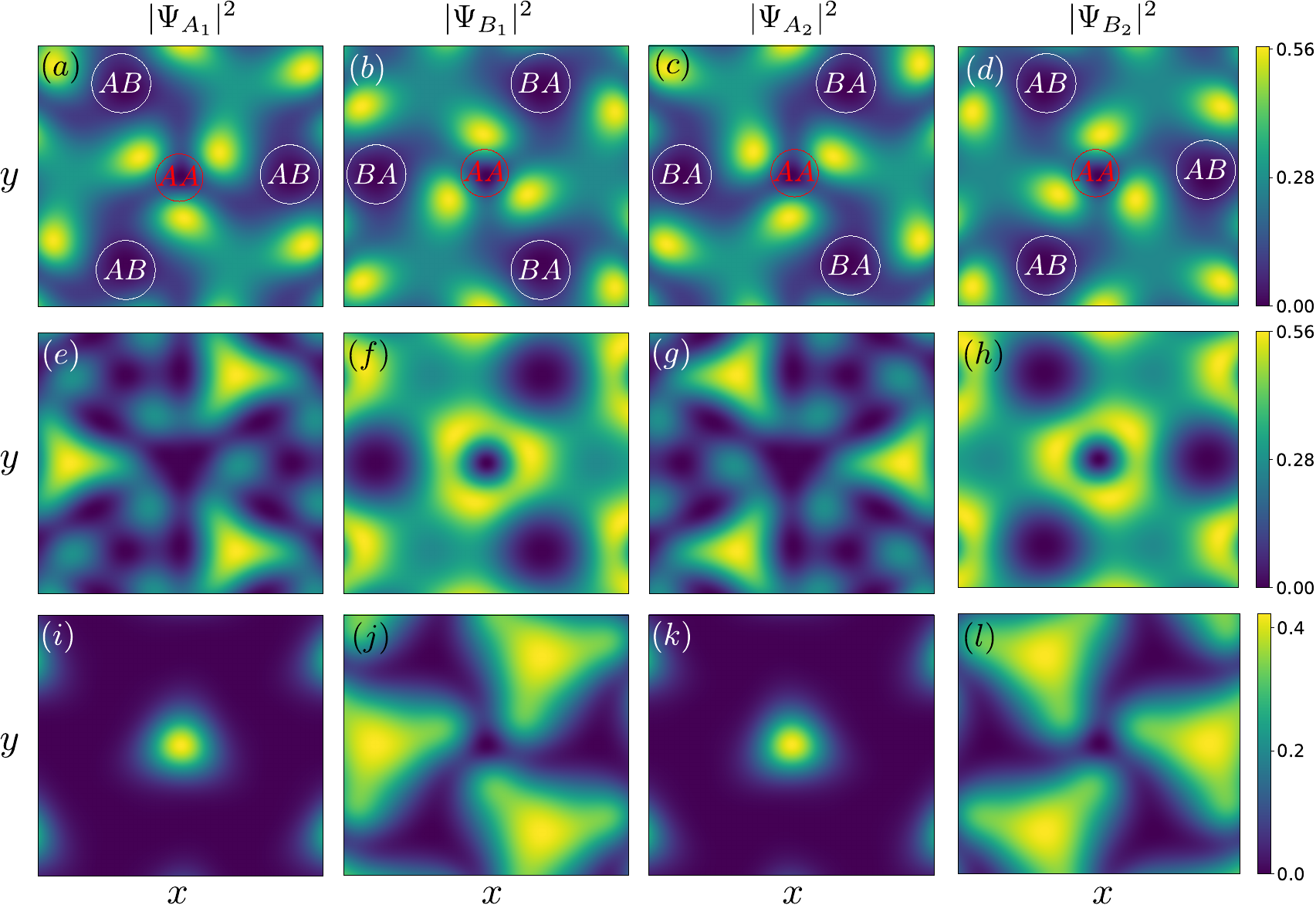}}
	\caption{Density plot for the squared amplitude of Bloch states are shown in the $x$ - $y$ plane for tBLG at a twist angle $\theta = 1.05^{o}$ near valley-$K$ and spin-up sector for the $p_{x} + ip_{y}$ superconducting state corresponding to the $\Gamma_{m}$-point of the first conduction band. The four columns respectively refer to the sublattice-A of layer-1 ($|\psi_{A_{1}}|^{2}$), sublattice-B of layer-1 ($|\psi_{B_{1}}|^{2}$), sublattice-A of layer-2 ($|\psi_{A_{2}}|^{2}$) and sublattice-B of layer-2 ($|\psi_{B_{2}}|^{2}$). The first row (\ie panels (a) - (d)), second row (\ie panels (e) - (h)) and third row (\ie panels (i) - (l)) correspond to $\mu = 1$ meV, $\mu = 10$ meV and $\mu = 20$ meV respectively choosing a fixed $\Delta_{\mathrm{sc}} = 10$ meV.
	}
	\label{Bloch_states}
\end{figure*}

Afterwards, we calculate the band gap between the first conduction and first valence bands throughout the mBZ and show the results in Fig.~\ref{tBLG_BZ_gap}. 
In Figs.~\ref{tBLG_BZ_gap}(a)-(c), we show the band gap for chemical potentials $\mu = 3$~meV, $\mu = 14.5$~meV, and $\mu = 20$~meV, respectively, for tBLG near the valley-$K$ sector and for spin-up states, with a superconducting order parameter $\Delta_{\mathrm{sc}} = 5$~meV and twist angle $\theta = 1.5^{\circ}$. These results correspond to the cut at $\theta = 1.5^{\circ}$ along the $\mu$ axis in Fig.~\ref{tBLG_Chern_phase}(a). A topological phase transition characterized by a change in the Chern number of three units (\ie $\mathcal{C}^{K}_{\uparrow}$ changes from $+2$ to $-1$) occurs at $\mu = 14.5$~meV. This transition can be attributed to a change in the chirality around three distinct gap-closing points within the mBZ that confirms the topological phase transition.

A similar gap-closing transition is illustrated in Figs.~\ref{tBLG_BZ_gap}(d)–(f) for tBLG near the valley-$K$ point and spin-up states, with $\theta = 1.05^{\circ}$ and $\Delta_{\mathrm{sc}} = 20$~meV. These panels correspond to the line at $\Delta_{\mathrm{sc}} = 20$~meV along varying $\mu$ in Fig.~\ref{tBLG_Chern_phase}(b). Here, we again observe a topological phase transition exhibiting a change in the Chern number by three ($\mathcal{C}^{K}_{\uparrow}: +2 \rightarrow -1$), where the gap closes at $\mu = 6.3$~meV at three distinct points near the $\Gamma_{m}$ point of the mBZ. We note that the different topological phases manifest distinct gap structure around the high symmetry points.

Finally, we investigate the spatial profiles of the Bloch-state amplitudes in real space, illustrated in Fig.~\ref{Bloch_states}, for layer-1 and layer-2 with sublattices A and B, focusing on the flat conduction band near the $\Gamma_{m}$ point (see Eq.~(\ref{Eq:Bloch_state})). This analysis helps us to understand the evolution of electronic states across different topological phases. The upper panel (Figs.~\ref{Bloch_states}(a)–(d)) exhibits results for $\Delta_{\mathrm{sc}} = 10$~meV and $\mu = 1$~meV at $\theta = 1.05^{\circ}$, corresponding to the $\mathcal{C}^{K}_{\uparrow} = -1$ phase in Fig.~\ref{tBLG_Chern_phase}(a). The middle panel (Figs.~\ref{Bloch_states}(e)–(h)) displays data for $\Delta_{\mathrm{sc}} = 10$~meV and $\mu = 10$~meV at $\theta = 1.05^{\circ}$, representing the topologically trivial phase with $\mathcal{C}^{K}_{\uparrow} = 0$. The lower panel (Figs.~\ref{Bloch_states}(i)–(l)) corresponds to $\Delta_{\mathrm{sc}} = 10$~meV and $\mu = 20$~meV at the same twist angle, capturing the topological phase with $\mathcal{C}^{K}_{\uparrow} = +2$. As can be seen from the figures, 
the different topological phases exhibit distinct Bloch localization profiles. Note that, in the trivial phase, the Bloch-state amplitudes indicate less localization than the topological phase.

%~~~~~~~~~~~~~~~~~~~~~~~~~~~~~~~~~~~~~~~~~~~~~~~~~~~~~~~~~
%~~~~~~~~~~~~~~~~~~~~~~~~~~~~~~~~~~~~~~~~~~~~~~~~~~~~~~~~~
%======================================================
\section{Topological superconducting properties of tDBLG}\label{Sec:V}
%======================================================

In this section, we explore the topological properties of tDBLG in both stacking configurations--AB-AB and AB-BA with chiral $p_{x} + ip_{y}$ superconducting pairing within each layer. We contrast the two stacking variants qualitatively and present a comparative analysis of their behavior alongside the results previously obtained for tBLG.

%~~~~~~~~~~~~~~~~~~~~~~~~~~~~~~~~~~~~~~~~~~~~~~~~~~~~~~~~~
%~~~~~~~~~~~~~~~~~~~~~~~~~~~~~~~~~~~~~~~~~~~~~~~~~~~~~~~~~
\subsection{Chern number}
%---------------------------------------------------------
%---------------------------------------------------------
\begin{figure}[h]
	\centering
	\subfigure{\includegraphics[width=0.5\textwidth]{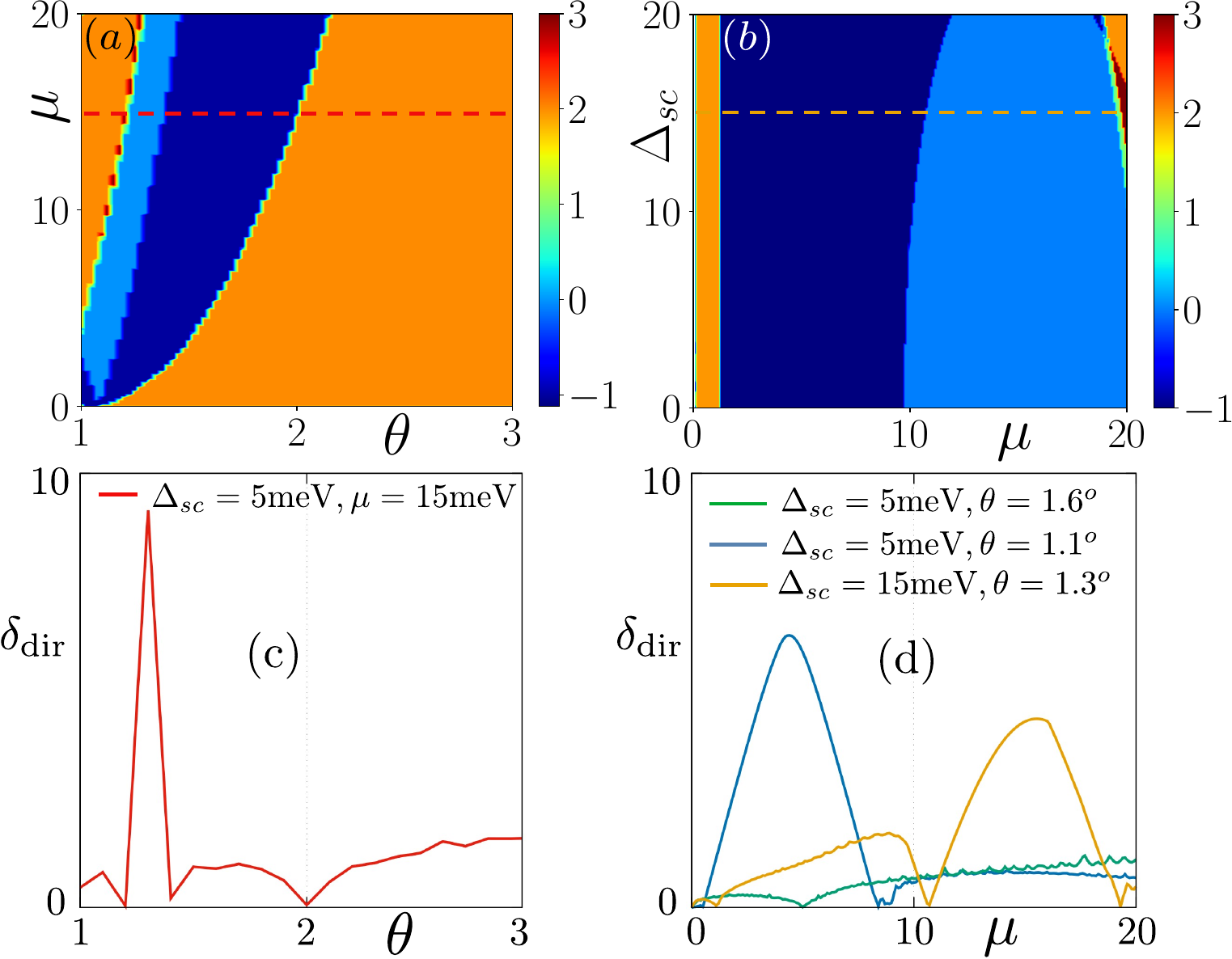}}
	\caption{Density plots and line plots are demonstrated respectively for the topological phase diagram and direct band gap in AB-AB stacked tDBLG with chiral $p_{x} + ip_{y}$ superconductivity around valley-$K$ and spin-up. In panel (a), Chern number ($\mathcal{C}^{K}_{\uparrow}$) is shown in the twist angle ($\theta$) and chemical potential ($\mu$) plane for superconducting order parameter ($\Delta_{\mathrm{sc}} = 5$ meV). In panel (b), the same is shown in the plane of chemical potential ($\mu$) and superconducting order parameter ($\Delta_{\mathrm{sc}}$) for the twist angle $\theta = 1.3^{o}$. 	
	Direct band gap ($\delta_{\text{dir}}$ in meV) is shown as a function of twist angle ($\theta$) and chemical potential ($\mu$) respectively in panels 
	(c) and in (d) for the same system with different values of twist angle ($\theta$), chemical potential ($\mu$) and superconducting order parameter ($\Delta_{\mathrm{sc}}$), as mentioned in the figures. 	 
	}
	\label{tDBLG(AB-AB)_Chern}
\end{figure}
%~~~~~~~~~~~~~~~~~~~~~~~~~~~~~~~~~~~~~~~~~~~~~~~~~~~~~~~~~~
%~~~~~~~~~~~~~~~~~~~~~~~~~~~~~~~~~~~~~~~~~~~~~~~~~~~~~~~~~~

%~~~~~~~~~~~~~~~~~~~~~~~~~~~~~~~~~~~~~~~~~~~~~~~~~~~~~~~~~~~~~~~~
%~~~~~~~~~~~~~~~~~~~~~~~~~~~~~~~~~~~~~~~~~~~~~~~~~~~~~~~~~~~~~~~~
\begin{figure}[h]
	\centering
	\subfigure{\includegraphics[width=0.5\textwidth]{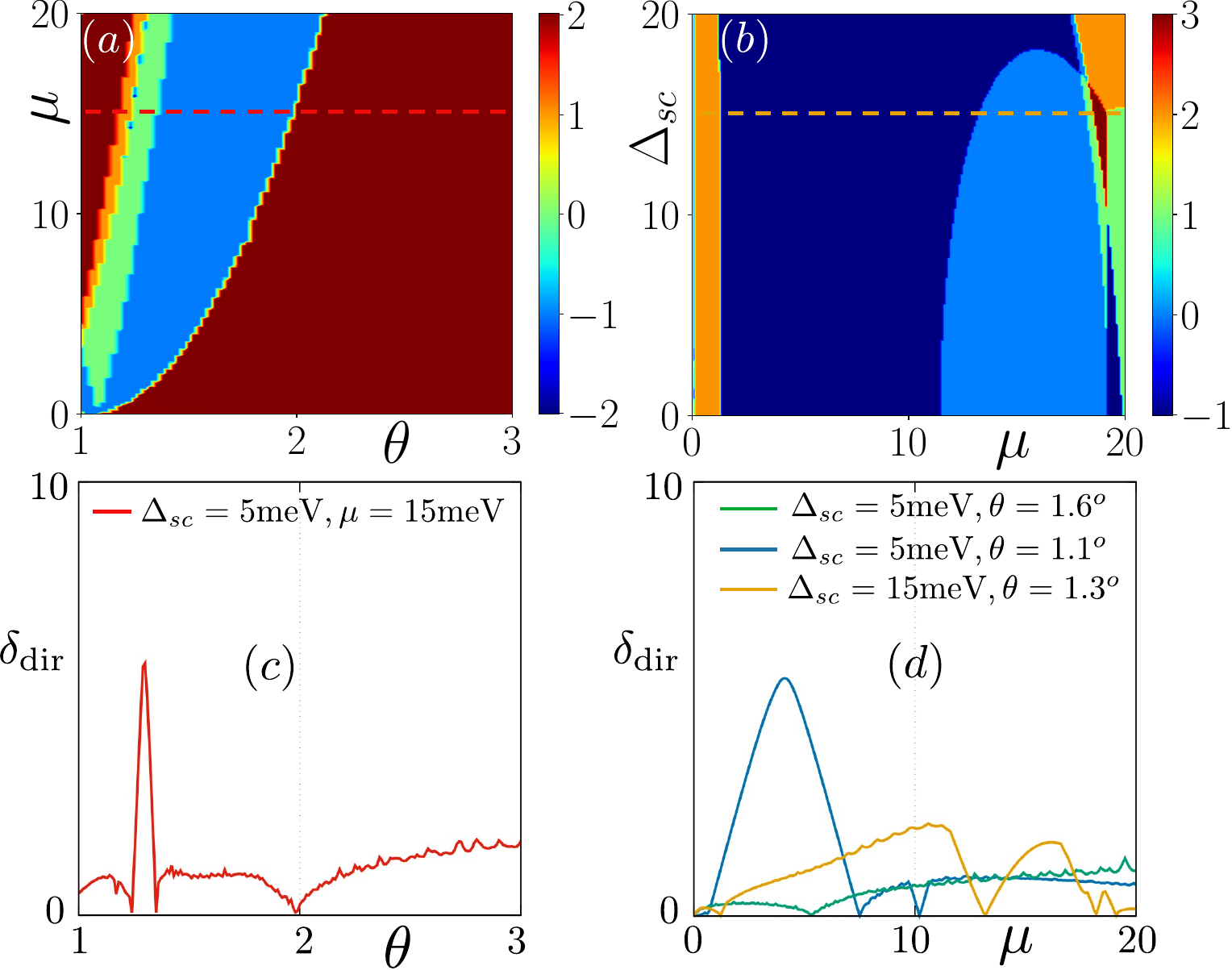}}
	\caption{Density plots and line plots are depicted respectively for the topological phase diagram and direct band gap in AB-BA stacked tDBLG with chiral $p_{x} + ip_{y}$ superconductivity around valley-$K$ and spin-up. In panel (a), Chern number ($\mathcal{C}^{K}_{\uparrow}$) is shown in the twist angle ($\theta$) and chemical potential ($\mu$) plane for superconducting order parameter ($\Delta_{\mathrm{sc}} = 5$ meV). In panel (b) the same is shown in
	the plane of chemical potential ($\mu$) and superconducting gap ($\Delta_{\mathrm{sc}}$) at the twist angle $\theta = 1.3^{o}$. Direct band gap ($\delta_{\text{dir}}$ in meV) is shown as a function of twist angle ($\theta$) and chemical potential ($\mu$) respectively in panels 
	(c) and in (d) for the same system with different values of twist angle ($\theta$), chemical potential ($\mu$) and superconducting order parameter ($\Delta_{\mathrm{sc}}$), as mentioned in the figures. 	 
	}
	\label{tDBLG(AB-BA)_Chern}
\end{figure}
%-------------------------------------------------------------
%-------------------------------------------------------------

To elucidate the topological properties of chiral $p_{x} + ip_{y}$ superconducting tDBLG systems, we compute the Chern number $\mathcal{C}^{K}_{\uparrow}$ around the valley-$K$ point and for spin-up ($s_z = +1$). We explore this system across a range of parameter values, following the same methodology employed for tBLG in the preceding section.

First, we consider the $(\theta, \mu)$-plane at a fixed superconducting pairing $\Delta_{\mathrm{sc}} = 5\,$meV and the features of Chern number $\mathcal{C}^{K}_{\uparrow}$ is shown in Fig.~\ref{tDBLG(AB-AB)_Chern}(a) and Fig.~\ref{tDBLG(AB-BA)_Chern}(a) for the AB-AB and AB-BA stacking configurations, respectively. Varying $\theta$ from $1^{\circ}$ to $3^{\circ}$, we find that at smaller twist angles a variety of topological phases appear (\eg \ $\mathcal{C}^{K}_{\uparrow} = -1$ and $+2$) over a wider parameter range than was seen for tBLG. A narrow stripe of the trivial phase (i.e.\ $\mathcal{C}^{K}_{\uparrow} = 0$) is also present in both tDBLG stackings. At larger twist angles, only the $\mathcal{C}^{K}_{\uparrow} = +2$ phase persists in
case of both AB-AB and AB-BA configurations. This behaviour can be explained in the similar way as for tBLG: as $\theta$ increases the flat-band bandwidth broadens while the chemical potential remains comparatively unchanged.

Then, we examine the same in $(\mu, \Delta_{\mathrm{sc}})$-plane at $\theta = 1.3^{\circ}$. The corresponding results are shown in Fig.~\ref{tDBLG(AB-AB)_Chern}(b) and Fig.~\ref{tDBLG(AB-BA)_Chern}(b) for the AB-AB and AB-BA stackings, respectively. In these density plots we identify several distinct topological phases including $\mathcal{C}^{K}_{\uparrow} = -1$, $+1$, $+2$, and $+3$. Although, a significant region remains topologically trivial (i.e.\ $\mathcal{C}^{K}_{\uparrow} = 0$). Both AB-AB and AB-BA stacked tDBLG qualitatively follow each other in terms of Chern number as well as gap closing profiles. However, the boundaries of these phases differ between the two stacking variants; there are parameter regime in which the AB-AB configuration is trivial, while AB-BA is topologically nontrivial. We further observe that at two limiting values of $\mu$, topological regions emerge. From the band perspective, for small $\mu$ the flat bands lie close to each other; as $\mu$ increases the separation between the flat bands grows, and near $\mu \approx 18\,$meV the flat bands again approach each other (near $\Gamma_{m}$ of the mBZ), leading to multiple topological phase transitions.

Finally, comparing tDBLG with tBLG, we note that while tBLG hosts a relatively extended trivial region in the $(\mu, \Delta_{\mathrm{sc}})$ parameter space, 
this trivial region becomes substantially reduced in tDBLG--especially for the AB-BA stacking case. Moreover, a new topological phase with $\mathcal{C}^{K}_{\uparrow} = +3$ appears in tDBLG, and the overall phase diagrams differ from those of tBLG. In a subsequent section, we discuss how the inclusion of trigonal-warping affects the topological phase diagram of tDBLG.

%~~~~~~~~~~~~~~~~~~~~~~~~~~~~~~~~~~~~~~~~~~~~~~~~~~~~~~~~~
%~~~~~~~~~~~~~~~~~~~~~~~~~~~~~~~~~~~~~~~~~~~~~~~~~~~~~~~~
%----------------------------------------------------
\begin{figure*}[t]
	\centering
	\subfigure{\includegraphics[width=0.95\textwidth]{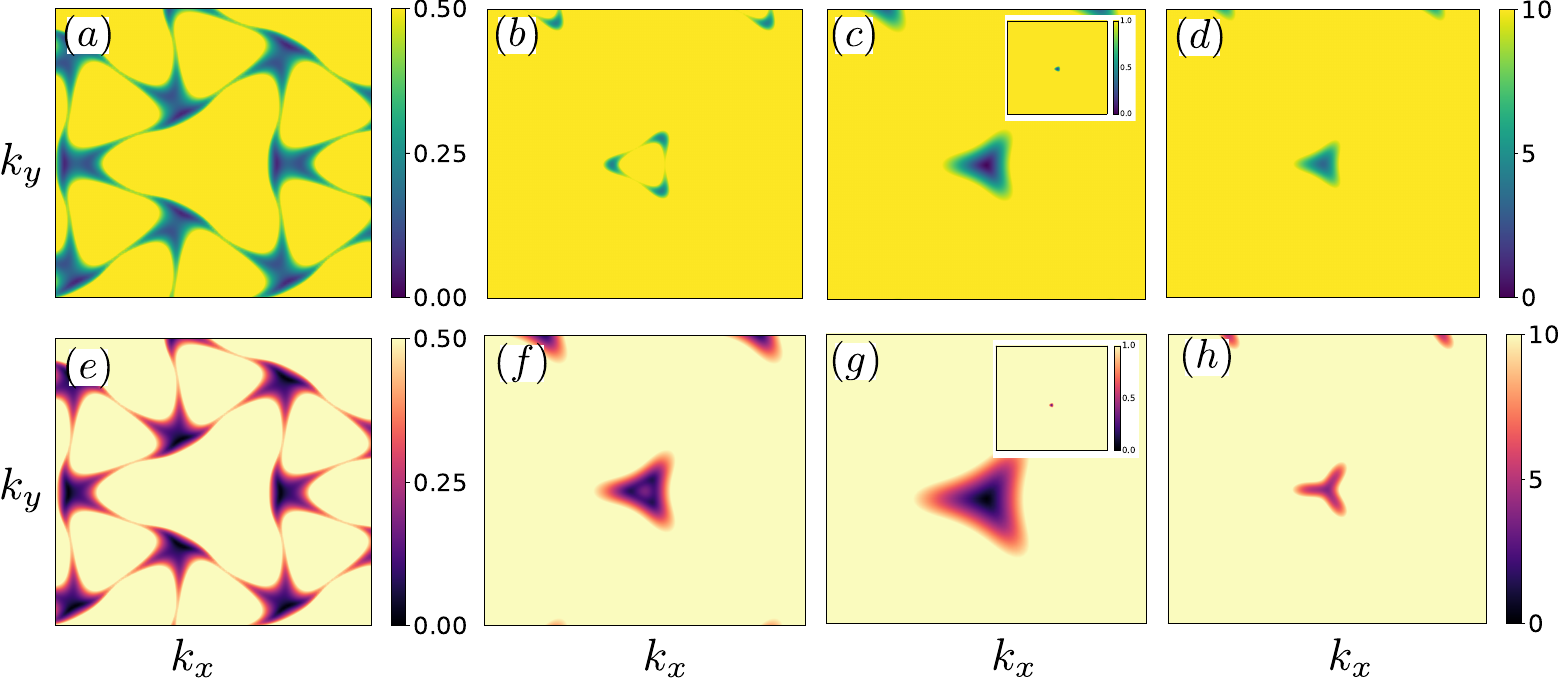}}
	\caption{(a) Density plot for the squared amplitude of Bloch states are depicted in the $x$ - $y$ plane for tDBLGs at a twist angle $\theta = 1.3^{o}$ near valley-$K$ and with the spin-up $p_{x} + ip_{y}$ superconducting order. These states belong to the $\Gamma_{m}$-point of the first conduction band. The four columns respectively refer to the sublattice-A of layer-1 ($|\psi_{A_{1}}|^{2}$), sublattice-B of layer-1 ($|\psi_{B_{1}}|^{2}$), sublattice-A of layer-4 ($|\psi_{A_{4}}|^{2}$) and sublattice-B of layer-4 ($|\psi_{B_{4}}|^{2}$).  . The first row (i.e., panel (a) - (d)), second row (i.e., panel (e) - (h)) and third row (i.e., panel (i) - (l))  correspond to $\mu = 0.5$ meV, $\mu = 5$ meV, $\mu = 15$ meV respectively, for a fixed $\Delta_{\mathrm{sc}} = 10$ meV in AB-AB tDBLG.}
	\label{tDBLG_BZ_gap1}
\end{figure*}

%---------------------------------------------------
%----------------------------------------------------
%--------------------------------------------
\subsection{Gap closing transitions}
%--------------------------------------------

To further understand the nature of topological phase transitions in tDBLGs with $p_{x} + ip_{y}$  pairing, we again compute the direct band gap defined in Eq.~(\ref{Eq:direct_band_gap}). We first focus on the AB-AB stacking, and subsequently discuss the AB-BA configuration.

In Fig.~\ref{tDBLG(AB-AB)_Chern}(c), we show the direct band gap $\delta_{\rm dir}$ for the AB-AB stacked tDBLG as a function of the twist angle $\theta$, for a fixed chemical potential $\mu = 15\,$meV and superconducting order parameter $\Delta_{\mathrm{sc}} = 5\,$meV. The corresponding dashed line in the phase diagram, indicated in Fig.~\ref{tDBLG(AB-AB)_Chern}(a), reveals three topological phase transitions: from $\mathcal{C}^{K}_{\uparrow} = +2$ to $0$, then $0$ to $-1$, and finally $-1$ to $+2$. Each of these transitions coincides with a closing of the direct band gap (red curve), occurring at $\theta = 1.2^{\circ}$, $1.6^{\circ}$ and $2.0^{\circ}$, respectively.

%~~~~~~~~~~~~~~~~~~~~~~~~~~~~~~~~~~~~~~~~~~~~~~~~~~~~~~~~~~~~~~~~~~~~~~~~~~~~~~~~~
%~~~~~~~~~~~~~~~~~~~~~~~~~~~~~~~~~~~~~~~~~~~~~~~~~~~~~~~~~~~~~~~~~~~~~~~~~~~~~~~~~
\begin{figure}[t]
	\centering
	\subfigure{\includegraphics[width=0.5\textwidth]{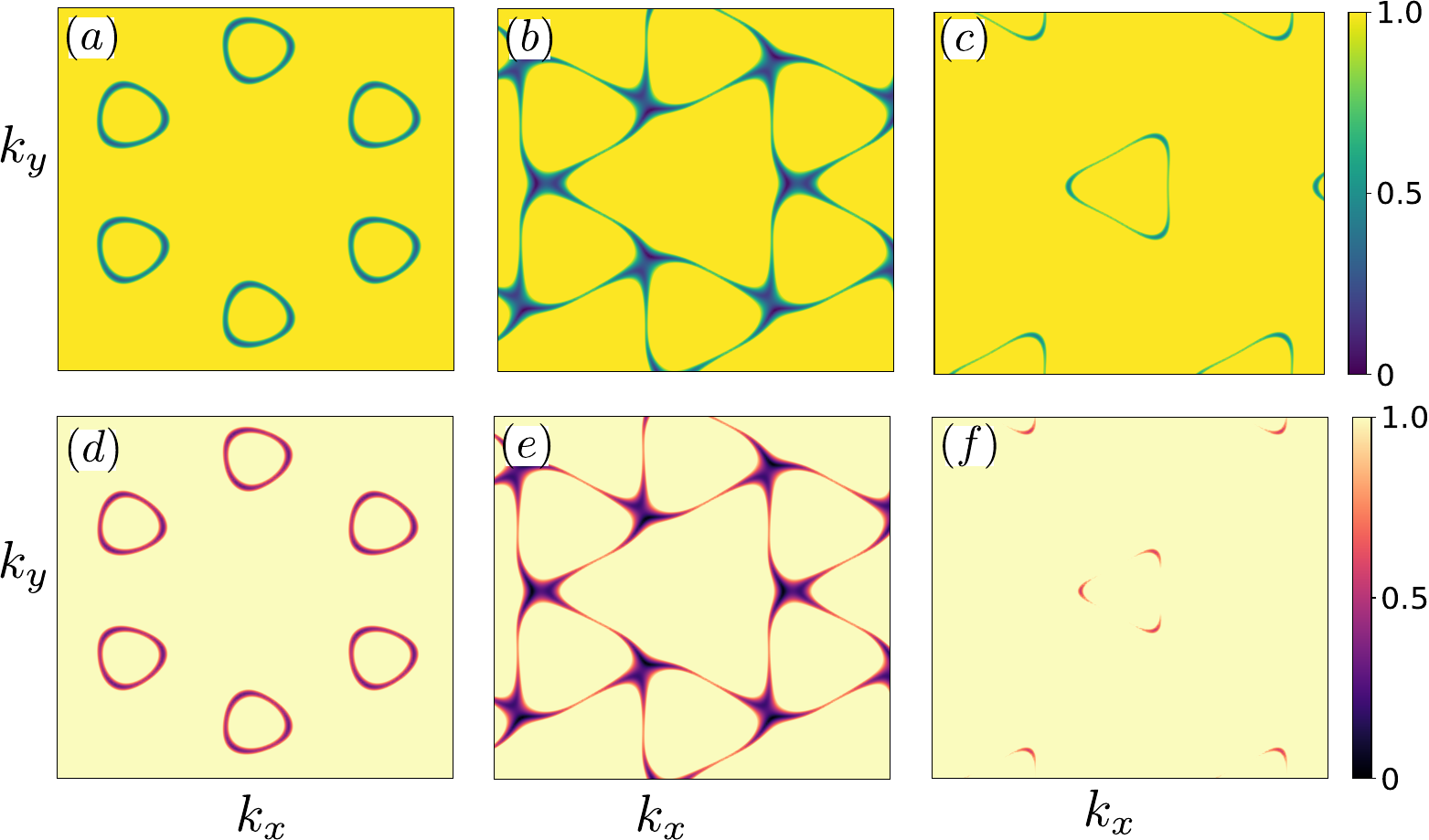}}
	\caption{Density plot for the band gap (in meV) are shown in the $k_{x} - k_{y}$ plane for tDBLG with $p_{x} + ip_{y}$ superconducting pairing and at a twist angle ($\theta = 1.6^{o}$) near valley-$K$ and spin-up. Panels (a) - (c) correspond to $\mu = 2$ meV, $\mu = 5$ meV, and $\mu = 15$ meV respectively for AB-AB stacked tDBLG at a fixed $\Delta_{\mathrm{sc}} = 5$ meV. Similarly, panels (d) - (f) correspond to $\mu = 2$ meV, $\mu = 5.3$ meV, and $\mu = 20$ meV respectively for AB-BA configured tDBLG at a fixed $\Delta_{\mathrm{sc}} = 5$ meV. 
	}
	\label{tDBLG_BZ_gap2}
\end{figure}
%~~~~~~~~~~~~~~~~~~~~~~~~~~~~~~~~~~~~~~~~~~~~~~~~~~~~~~~~~~~~~~~~~~~~~~~~~~~~~~~~~~~
%~~~~~~~~~~~~~~~~~~~~~~~~~~~~~~~~~~~~~~~~~~~~~~~~~~~~~~~~~~~~~~~~~~~~~~~~~~~~~~~~~~~

In Fig.~\ref{tDBLG(AB-AB)_Chern}(d), we showcase $\delta_{\rm dir}$ with respect to the chemical potential $\mu$. The green line corresponds to the line cut at $\theta = 1.6^{\circ}$ with $\Delta_{\mathrm{sc}} = 5\,$meV in Fig.~\ref{tDBLG(AB-AB)_Chern}(a), and exhibits a single topological transition, namely from $\mathcal{C}^{K}_{\uparrow} = +2$ to $-1$. The blue line corresponds to $\theta = 1.1^{\circ}$ with $\Delta_{\mathrm{sc}} = 5\,$meV (same phase diagram), and exhibits multiple gap closings that mark successive topological transitions. The yellow line corresponds to the dashed line at $\theta = 1.3^{\circ}$ with $\Delta_{\mathrm{sc}} = 15\,$meV in Fig.~\ref{tDBLG(AB-AB)_Chern}(b), and reflects three transitions: from $\mathcal{C}^{K}_{\uparrow} = +2$ to $-1$, then $-1$ to $0$, and finally $0$ to $+3$, occurring at $\mu = 1\,$meV, $10.7\,$meV and $19.3\,$meV respectively.

Turning to the AB-BA stacking, Fig.~\ref{tDBLG(AB-BA)_Chern} manifests the direct band gap for the same set of parameter cuts as mentioned above. In Figs.~\ref{tDBLG(AB-BA)_Chern}(c) and (d), we display the bulk gap as a function of $\theta$ and $\mu$ respectively. Again, we find that each topological phase transition is associated with a vanishing $\delta_{\rm dir}$. However, the precise parameter values at which the transitions occur differ from those of the
AB-AB stacking.

We now discuss the band-gap between the first conduction and first valence bands across the mBZ. In Figs.~\ref{tDBLG_BZ_gap1}(a)-(d), the gap is shown for the AB-AB stacking of tDBLG at $\Delta_{\mathrm{sc}} = 10\,$meV and $\theta = 1.3^{\circ}$, with chemical potentials set to $\mu = 1.2\,$meV, $6\,$meV, $10\,$meV, and $12\,$meV, respectively. In Figs.~\ref{tDBLG_BZ_gap1}(e)–(h) the corresponding results are presented for the AB-BA stacking, again at $\Delta_{\mathrm{sc}} = 10\,$meV and $\theta = 1.3^{\circ}$, with $\mu = 1.2\,$meV, $10\,$meV, $12\,$meV, and $15\,$meV respectively. Note that, different topological phases exhibit distinct gap profile. For AB-AB stacking, Fig.~\ref{tDBLG_BZ_gap1}(a) shows three distinct gap‐closing points within the mBZ, while Fig.~\ref{tDBLG_BZ_gap1}(c) displays a single gap closing event near the $\Gamma_{m}$-point. Accordingly, along the $\Delta_{\mathrm{sc}} = 10\,$meV cut in Fig.~\ref{tDBLG(AB-AB)_Chern}(b), we observe corresponding changes in the Chern number of magnitude 3 and 1 respectively. Similar gap‐closing transitions can be seen for the AB-BA stacking 
(see Figs.~\ref{tDBLG_BZ_gap1}(e) and (g)), and the corresponding match with the change in Chern number can be found in Fig.~\ref{tDBLG(AB-BA)_Chern}(b).  

In Figs.~\ref{tDBLG_BZ_gap2}(a)–(c), we show the band gap for AB-AB stacking at $\Delta_{\mathrm{sc}} = 5\,$meV and $\theta = 1.6^{\circ}$, with $\mu = 2\,$meV, $5\,$meV and $15\,$meV respectively. This corresponds to the line cut at $\theta = 1.6^{\circ}$ with $\Delta_{\mathrm{sc}} = 5\,$meV in Fig.~\ref{tDBLG(AB-AB)_Chern}(a). It can be seen that at $\mu = 5\,$meV three gap‐closing points appear in the mBZ, associated with a change in the Chern number. Similarly, for the AB-BA stacking (Figs.~\ref{tDBLG_BZ_gap2}(d)–(f)), we depict the band gap for $\mu = 2\,$meV, $5.3\,$meV and $20\,$meV (with $\Delta_{\mathrm{sc}} = 5\,$meV and $\theta = 1.6^{\circ}$). Upon careful observation, one finds that, just as in the AB-AB case, for AB-BA too the number of gap-closing points inside the mBZ corresponds to the change in the Chern number as manifested in Fig.~\ref{tDBLG(AB-BA)_Chern}(a).

\begin{figure*}[t]
	\centering
	\subfigure{\includegraphics[width=0.85\textwidth]{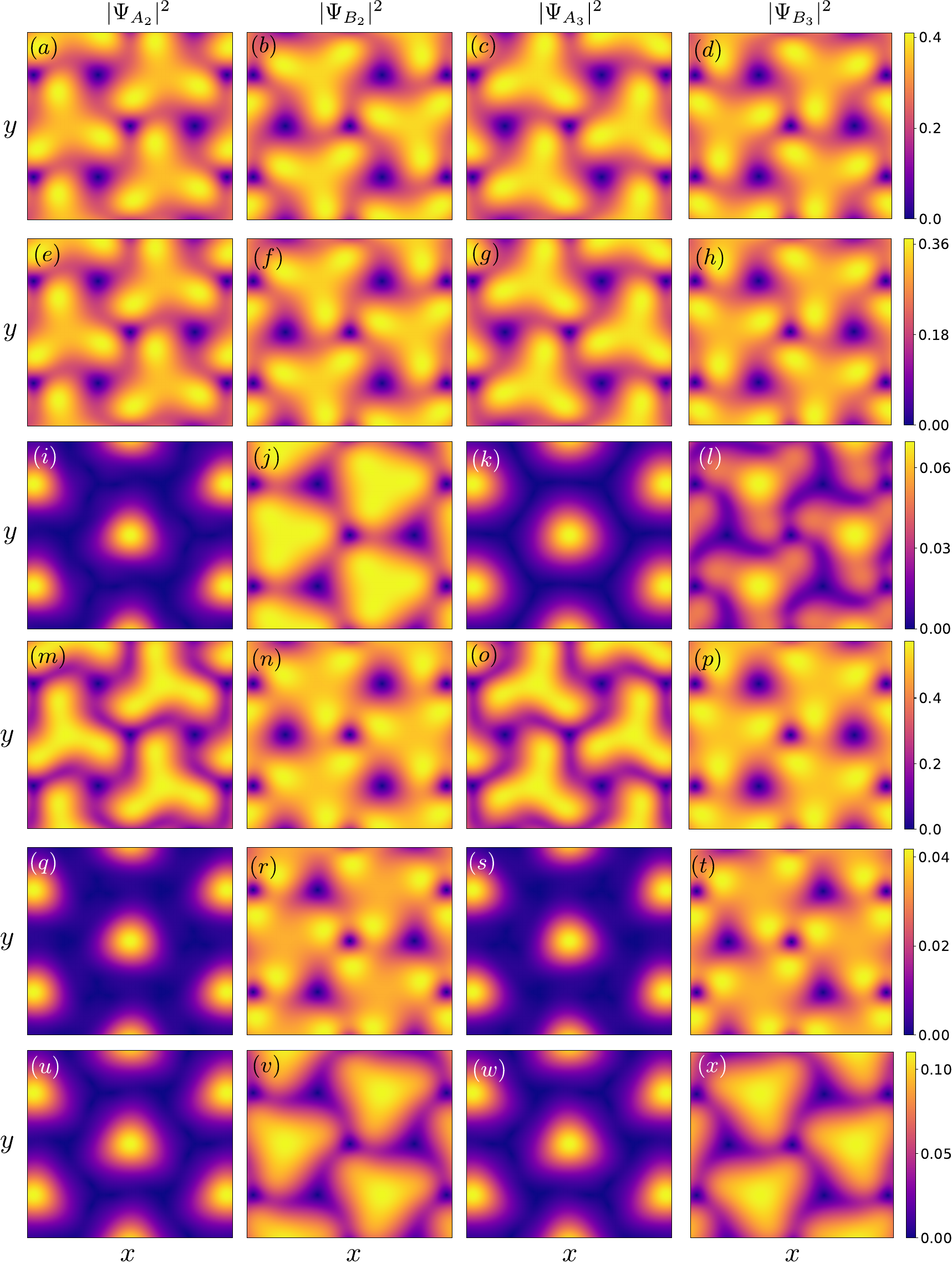}}
	\caption{Density plot for the squared amplitude of Bloch states are depicted in the $x$ - $y$ plane for tDBLGs incorporating the $p_{x} + ip_{y}$ superconducting order. These states belong to the $\Gamma_{m}$-point of the first conduction band. The four columns respectively refer to the sublattice-A of layer-2 ($|\psi_{A_{2}}|^{2}$), sublattice-B of layer-2 ($|\psi_{B_{2}}|^{2}$), sublattice-A of layer-3 ($|\psi_{A_{3}}|^{2}$) and sublattice-B of layer-3 ($|\psi_{B_{3}}|^{2}$). Panels (a) - (l) and (m)- (x) are shown respectively for AB-AB and AB-BA tDBLG at a twist angle $\theta = 1.3^{o}$ near valley-$K$ and spin-up. The first row (\ie panel (a) - (d)), second row (\ie panel (e) - (h)) and third row (\ie panel (i) - (l))  correspond to $\mu = 0.5$ meV, $\mu = 5$ meV, $\mu = 15$ meV respectively, for a fixed $\Delta_{\mathrm{sc}} = 10$ meV in AB-AB tDBLG. Also, the fourth row (\ie panel (m) - (p)), fifth row (\ie panel (q) - (t)) and sixth row (\ie panel (u) - (x)) correspond to $\mu = 0.5$ meV, $\mu = 5$ meV, $\mu = 15$ meV respectively, for a fixed $\Delta_{\mathrm{sc}} = 10$ meV in case of AB-BA tDBLG.  
	}
	\label{AB_AB_Bloch}
\end{figure*}

Finally, we compute the spatial profiles of the Bloch-state amplitudes in real space for both the AB-AB and AB-BA stacked tDBLG systems (see Fig.~\ref{AB_AB_Bloch}). We focus on layer-2 and layer-3 with sublattices A and B (as they directly participate in the twisting mechanism), specifically for the conduction band near the mBZ centre $\Gamma_{m}$. 
In Fig.~\ref{AB_AB_Bloch}, the upper panels (a)–(d) correspond to parameters $\Delta_{\mathrm{sc}} = 10\,$meV and $\mu = 1\,$meV at a twist angle $\theta = 1.3^{\circ}$, which lies in the topological phase $\mathcal{C}^{K}_{\uparrow} = +2$ (see Fig.~\ref{tDBLG(AB-AB)_Chern}(b) and Fig.~\ref{tDBLG(AB-BA)_Chern}(b)). The middle panels (e)–(h) exhibit data for $\Delta_{\mathrm{sc}} = 10\,$meV and $\mu = 10\,$meV 
at the same twist angle, corresponding to the phase $\mathcal{C}^{K}_{\uparrow} = -1$. 
The lower panels (i)–(l) are for $\Delta_{\mathrm{sc}} = 10\,$meV and 
$\mu = 15\,$meV with $\theta = 1.3^{\circ}$, and represent the trivial phase $\mathcal{C}^{K}_{\uparrow} = 0$. Similarly, panels (m)-(x) in Fig.~\ref{AB_AB_Bloch} refer to the spatial profiles of the Bloch-state amplitudes for AB-BA stacked tDBLG. In Appendix.~\ref{AppE}, we also present the Bloch state profiles for the outer layers of tDBLG (\ie layer-1 and layer-4). 

%-------------------------------------------------------------------
%-------------------------------------------------------------------
%AB_AB%
%----------------------------------------------------------------
%----------------------------------------------------------------
%AB_BA%

%======================================================
\section{Effects of trigonal warping in tDBLG}\label{Sec:VI}
%======================================================
%~~~~~~~~~~~~~~~~~~~~~~~~~~~~~~~~~~~~~~~~~~~~~~~~~~~~~~~
%~~~~~~~~~~~~~~~~~~~~~~~~~~~~~~~~~~~~~~~~~~~~~~~~~~~~~~~
\begin{figure}[h]
	\centering
	\subfigure{\includegraphics[width=0.49\textwidth]{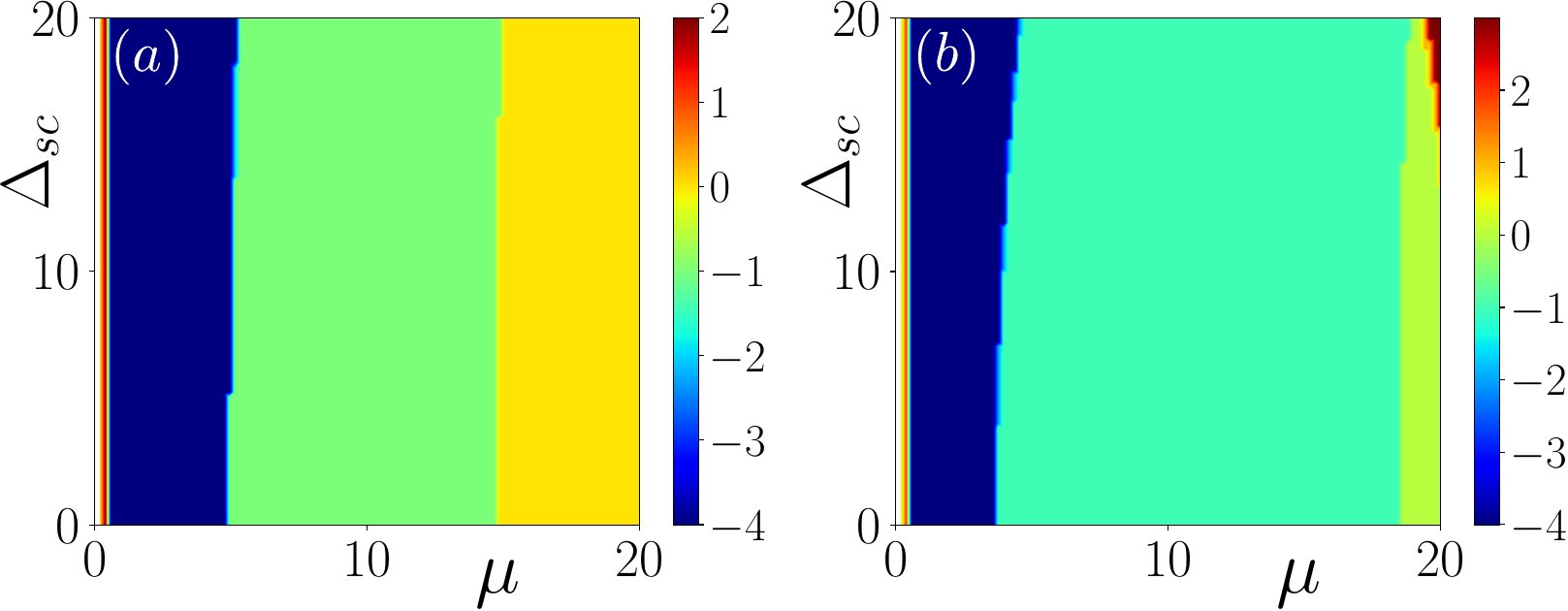}}
	\caption{Density plots corresponding to the topological phase diagram are displayed for tDBLG (in presence of trigonal warping) with chiral $p_{x} + ip_{y}$ superconductivity around valley-$K$ and spin-up. In panels (a) and (b), Chern number ($\mathcal{C}^{K}_{\uparrow}$) is shown in the chemical potential ($\mu$) and superconducting order parameter ($\Delta_{\mathrm{sc}}$) plane at the twist angle $\theta = 1.3^{o}$ for AB-AB and AB-BA tDBLG respectively.  
	}
	\label{tDBLGs_Chern_trigon_warp}
\end{figure}
%~~~~~~~~~~~~~~~~~~~~~~~~~~~~~~~~~~~~~~~~~~~~~~~~~~~~~~~~
%~~~~~~~~~~~~~~~~~~~~~~~~~~~~~~~~~~~~~~~~~~~~~~~~~~~~~~~~

In bilayer graphene, trigonal warping (\ie $\gamma_{3}$ in Eq.~(\ref{Eq:BLG_interlayer}) plays a crucial role in controlling the topological behaviour, as it directly modifies the Fermi surface topology and thereby the band-inversion mechanisms. In the following, we present the topological phase diagrams for both the AB-AB and AB-BA stacked tDBLG systems in the presence of trigonal warping. In Figs~\ref{tDBLGs_Chern_trigon_warp}(a) and (b) we display the Chern number $\mathcal{C}^{K}_{\uparrow}$ in the $(\mu,\Delta_{\mathrm{sc}})$ plane at a fixed twist angle $\theta = 1.3^{\circ}$, for the AB-AB and AB-BA stacked tDBLGs respectively. In this case, we observe multiple topological phases such as $\mathcal{C}^{K}_{\uparrow} = -4,\,-1,\,+2,\,+3$. It is evident that the inclusion of trigonal warping leads to a significant restructuring of the phase diagrams: a new topological phase with $\mathcal{C}^{K}_{\uparrow} = -4$ appears over an extended region of parameter space in both stackings. Another notable change is that, for the parameter ranges considered, the topologically non-trivial regions expand while the trivial phase region becomes narrower, compared to the diagrams without trigonal warping shown in Figs.~\ref{tDBLG(AB-AB)_Chern}(b) and \ref{tDBLG(AB-BA)_Chern}(b). In particular, for the AB-BA tDBLG (see Fig.~\ref{tDBLGs_Chern_trigon_warp}(b)), the entire phase diagram becomes topologically non-trivial, except for a narrow region near $\mu \approx 20\,$meV.

%======================================================
\section{Possible route to realize effective $p + ip$-pairing from $s$-wave superconductivity}\label{Sec:VII}
%======================================================
In this section, we discuss the possibility of realizing the $p+ip$-superconducting pairing as an effective pairing using $s$-wave superconductor~\cite{Dual_Sato,Dual_pritam}. Our model for tBLG comprises of monolayer graphene's pairing (intralayer) and we do not have any interlayer superconducting pairing. Hence, to keep things simple we investigate the monolayer graphene and fetch our conclusion regarding tBLG.

The BdG Hamiltonian for a single layer of graphene under the low energy approximation (near valley-$K$) is derived in Appendix~\ref{AppA} and can be written in the basis $\bigl( A_{k\uparrow}, B_{k\uparrow}, A_{k\downarrow}, B_{k\downarrow}, A^{\dagger}_{-k\uparrow}, B^{\dagger}_{-k\uparrow}, A^{\dagger}_{-k\downarrow}, B^{\dagger}_{-k\downarrow} \bigr)^{T}$ as,
\begin{align}
	h_{\text{BdG}}({\bf q})= 
	\left( \begin{array}{cc}
		h_{\text{SLG}}({\bf q}) & h_{\Delta}({\bf q})\\
		h^\dagger_{\Delta}({\bf q}) & -h_{\text{SLG}}({\bf q}) 
	\end{array}\right)\ ,
	\label{BdG_SLG_low1}
\end{align}	
where, 
\begin{align}
	h_{\text{SLG}}({\bf q}) = 
	\left( \begin{array}{cccc}
		\mu & v_{F} q_{+} & 0 & 0\\
		v_{F} q_{-} & \mu & 0 & 0\\ 
		0 & 0 & \mu & v_{F} q_{+}\\
		0 & 0 & v_{F} q_{-} & \mu \\ 
	\end{array}\right)\ ,
	\label{Eq:H:BLG1}
\end{align}	
and,
\begin{align}
	h_{\Delta}({\bf q}) = 
	\left( \begin{array}{cccc}
		0 & i \sqrt{\frac{3}{2}} \Delta_{sc} & 0 & 0\\
		-\frac{1}{2} i \sqrt{\frac{3}{2}} \Delta_{sc} q_{+} & 0 & 0 & 0 \\
		0 & 0 & 0 & \frac{1}{2} i \sqrt{\frac{3}{2}} \Delta_{sc} q_{-}\\
		0 & 0 & -i \sqrt{\frac{3}{2}} \Delta_{sc} & 0 
	\end{array}\right)\ .
	\label{SLG_pairing}
\end{align}	

Then, we introduce a unitary transformation $U$, to find a dual Hamiltonian corresponding to the BdG-Hamiltonian (Eq.~(\ref{BdG_SLG_low1})) as,
\begin{eqnarray}
	U = I_{4} \otimes S ,
\end{eqnarray}
with, $I_{4}$ is a 4-by-4 identity matrix and
\begin{eqnarray}
	S = \left(
	\begin{array}{cc}
		\frac{1}{\sqrt{2}} & \frac{1}{\sqrt{2}} \\
		-\frac{1}{\sqrt{2}} & \frac{1}{\sqrt{2}} \\	
	\end{array}
	\right)
\end{eqnarray}
Therefore, performing the dual transformation we obtain the dual Hamiltonian as,
\begin{eqnarray}
	H_{BdG}^{D} = U^{\dagger} H_{BdG} U\ .
\end{eqnarray}
Then, one can write the Hamiltonian with the block matrices as,
\begin{eqnarray}
	H_{BdG}^{D} = \left(
	\begin{array}{cc}
		H_{00}^{D} & H_{01}^{D} \\
		H_{10}^{D} & H_{11}^{D} \\	
	\end{array}
	\right)\ ,
\end{eqnarray}
where,
%\begin{widetext}
	\begin{align}
		H_{00}^{D} &=& \left(
		\begin{array}{cccc}
			0 & -\frac{i}{4}  \Delta (q_{-}+2) & 0 & 0 \\
			\frac{i}{4}  \Delta (q_{+}+2) & 0 & -\frac{i\Delta}{2} & 0 \\
			0 & \frac{i\Delta}{2} & 0 & -\frac{i}{4}  \Delta (q_{-}+2)\\
			0 & 0 & \frac{i}{4}  \Delta (q_{+}+2) & 0 \\
		\end{array}
		\right)\ ,
		\end{align}
		
		\begin{widetext}
		\begin{eqnarray}
		H_{01}^{D} &=& \left(
		\begin{array}{cccc}
			\mu  &- \frac{i}{4}  \Delta (q_{-}-2) +  v_F q_{+}  & 0 & 0 \\
			-\frac{i}{4} \Delta (q_{+}-2
			i)+v_F q_{-} & \mu  & 0 & 0 \\
			0 & 0 & \mu  & \frac{i}{4} \Delta (q_{-}-2))+v_F q_{+} \\
			0 & 0 & \frac{i}{4}  \Delta (q_{+}-2) + v_F q_{-} & \mu  \\
		\end{array}
		\right)\ ,\\ \ 
		H_{10}^{D} &=& \left(
		\begin{array}{cccc}
			\mu  & \frac{i}{4} \Delta (q_{-}+2i) + v_F q_{+} & 0 & 0\\
			\frac{1}{4} i \Delta (q_{+}-2) + v_F q_{-} & \mu  & 0 & 0\\
			0 & 0 & \mu  & v_F q_{+} - \frac{1}{4} i \Delta (q_{-}-2)\\
			0 & 0 & -i\frac{1}{4} \Delta (q_{+}-2 ) + v_F q_{-} & \mu \\
		\end{array}
		\right)\ ,\\ \ 
		H_{11}^{D} &=& \left(
		\begin{array}{cccc}
			0 & \frac{i}{4} \Delta (q_{-} + 2) & 0 & 0 \\
			-\frac{i}{4} \Delta (q_{+} + 2)) & 0 & \frac{i\Delta}{2} & 0 \\
			0 & -\frac{i\Delta}{2} & 0 & \frac{i}{4} \Delta (q_{-} + 2)) \\
			0 & 0 & -\frac{i}{4} \Delta (q_{+} + 2) & 0 \\
		\end{array}
		\right)\ .
	\end{eqnarray}
\end{widetext}

Here, $\sqrt{\frac{3}{2} }\Delta_{\mathrm{sc}}$ is denoted as $\Delta$. From the diagonal blocks (\ie $H_{00}^{D}$ and $H_{11}^{D}$) we can identify an effective magnetic field, 
$B = \frac{\Delta}{2} \sigma_{y} \otimes \tau_{0}$. Also, from the same $H_{00}^{D}$ and $H_{11}^{D}$ one can identify akin to RSOC for graphene (\ie $\frac{i \Delta}{4}$ at the (1,2)$^{\rm{th}}$ position in $H_{00}^{D}$). The Pauli matrices $\sigma$ and $\tau$ act on the spin and sublattice degrees of freedom respectively. On the other hand, the off diagonal blocks of $H_{00}^{D}$ and $H_{11}^{D}$ only exhibit effective spin-singlet $s$-wave pairings. In particular, the momentum independent term (like, $\mu$) are similar to the effective $s$-wave pairing, while momentum dependent terms can be attributed to the extended $s$-wave like pairing. 

We model our tBLG and tDBLG systems by stacking layers of graphene sheets with existing $p_{x} + ip_{y}$ superconducting pairing in them. Hence, ensuring each individual layer of graphene would inherit the essential components like conventional $s$-wave superconductivity, RSOC and magnetic field, then that system in a multilayer setup 
can possibly mimic our model Hamiltonian of $p_{x} + ip_{y}$ pairing.
%======================================================

%======================================================
\section{Summary and Conclusions}\label{Sec:VIII}
%======================================================

To summarize, in this article, we construct a model Hamiltonian for tBLG and tDBLG in presence of chiral $p_{x} + ip_{y}$-superconductiviting order. Starting from the Bistritzer-MacDonald model, we write the BdG Hamiltonians for tBLG and tDBLG. Initially, we discuss the electronic band properties of tBLG and tDBLG in presence of the unconventional superconducting pairing and chemical potential. In the latter part, topological superconducting properties are discussed for both tBLG and tDBLG. Considering the BdG-Hamiltonian, we calculate the Chern numbers to characterize the topological band properties of the systems. Since our model Hamiltonian is spin and valley degenerate, we only focus on a single valley and spin-up sector. Our findings for the topological phases are presented in two phase diagrams: one is shown in the ($\mu, \theta$) plane for a fixed value of the superconducting order parameter ($\Delta_{sc}$), the other phase diagram is depicted in the ($\Delta, \mu$) plane for a fixed value of the twist angle. While for both tBLG and tDBLG various topological phases with different Chern numbers (\eg $\mathcal{C}^{K}_{\uparrow} = -1, 0, 2, 3$) appear, however all the three systems (tBLG, AB-AB and AB-BA stacked tDBLG) manifest different phase boundaries. As a result, non-topological regions of one system becomes topological in other systems for the same parameter values.

In a latter part of the article, we also calculate the direct band gap to establish the observed topological phases and corresponding phase transitions. We present several lines with varying parameter values that crosses the topological phases, and identify the direct band gap closings at the phase transition points. Further, we calculate the band gap on the mBZ along few topologial phase transition lines to understand the gap closings on the Fermi surface and corresponding change in Chern number of the system. Finally, the evolution of squared amplitudes of the Bloch states for different sublattices and layers are shown for parameter values lying in different topological phases. Also, we briefly discuss the effect of trigonal warping on the topological phase diagram of AB-AB and AB-BA stacked tDBLGs. Additional interesting topological phase with $\mathcal{C}^{K}_{\uparrow} = -4$ arises for an extended range of parameter values for both the types of tDBLGs. Also, the non-topological region decreases, specially in AB-BA tDBLG case it almost turns into topological for almost all the parameter values. We end our discussion by providing a possible route to realize the  $p+ip$-superconductivity in tBLG as an effective pairing arising from a conventional $s$-wave superconductor in presence of RSOC and magnetic field. 

%Distinction from previous study:
Although the topological properties of chiral superconductivity in tBLG have been explored previously~\cite{topo_TSC_tBLG}, our work differs from that in a couple of important ways. The earlier study focused on a six-band model Hamiltonian limited to the flat-band regime around the magic angle, whereas we employ the low-energy continuum model proposed by Bistritzer and MacDonald~\cite{MacDonald-tBLG} to construct the full BdG Hamiltonian for tBLG as well as tDBLG, allowing us to tune the twist angle and investigate behavior both at and away from the magic angle. Additionally, the pairing structure is treated differently: instead of defining pairing on the emergent moiré lattice, we begin with pairing functions for single-layer graphene, apply a low-energy approximation, and incorporate them into the BdG framework, leading to a distinct microscopic picture of superconductivity in twisted case.

%Possible experimental realization
While tBLG itself exhibits unconventional superconductivity near the magic angle, the nature of such pairing is still unclear. In several theoretical studies $d+id$, $p+ip$-like pairing and even coexistence of both is predicted~\cite{DMFT_TSC_tBLG,RG_TSC_tBLG,RG_TSC_tDBLG,MFT_TSC_tBLG,QMC_TSC_tBLG,BdG_TSC_tBLG}. On the other hand, very recently, signatures of chiral topological superconductivity in rhombohedral graphene~\cite{Han2025} have been reported. This arises from a spin- and valley-polarized quarter-metal state in which only a single valley contributes to the Fermi surface. In contrast, the inter-valley pairing considered in our work is qualitatively distinct from the intravalley pairing mechanism operative in rhombohedral graphene~\cite{RH_graphene_theory1,RH_graphene_theory2}. Nevertheless, these developments may offer valuable insights toward engineering topological superconductivity in twisted and multilayer graphene based systems in the near future.

%Bloch state related clarification:
Finally note that, although we present the squared amplitude of Bloch states for different topological phases and observe distinct real-space profiles for some phases, however, several phases remain indistinguishable in this regard. Therefore, it is not generally true that Bloch state amplitude profiles can always differentiate between topological and trivial phases. Consequently, one should not rely solely on such local probes; instead, a proper topological characterization requires the detailed analysis of an appropriate invariant, such as the Chern number.

%======================================================
\subsection*{Acknowledgments}
%======================================================
T.N. thanks Snehasish Nandy, Sanjib Das and Lucas Baldo for useful discussions on tBLG. T.N. acknowledges the NFSG
“NFSG/HYD/2023/H0911” from BITS Pilani. K.B. and A.S. acknowledge SAMKHYA: High-Performance Computing Facility provided by Institute of Physics, Bhubaneswar, 
and the two workstations provided by the Institute of Physics, Bhubaneswar from the DAE APEX project for numerical computations.

%======================================================

%======================================================
\subsection*{Data Availability Statement}
%======================================================
The datasets generated and analyzed during the current study are available from the corresponding author upon reasonable request.

%                    Appendix
%%%%%%%%%%%%%%%%%%%%%%%%%%%%%%%%%%%%%%%%%%%%%%%%%%%%%%%%%%%%%%%%%%%%%%%%%%%%%%%%%%%%%
\appendix
%~~~~~~~~~~~~~~~~~~~~~~~~~~~~~~~~~~~~~~~~~~~~~~~~~~~~~~~~~~~~~~~~~~~~~~~~~~~~~~~~~~~~~~~~~~~~~~~~~~~~
\section{DERIVATION OF THE FORM FACTORS FOR $p_{x} + ip_{y}$ PAIRING IN GRAPHENE}
\label{AppA}
%~~~~~~~~~~~~~~~~~~~~~~~~~~~~~~~~~~~~~~~~~~~~~~~~~~~~~~~~~~~~~~~~~~~~~~~~~~~~~~~~~~~~~~~~~~~~~~~~~~~~

%superconducting pairing structure:
In order to discuss the pairing structure in details, we start from the real space mean-field Hamiltonian~\cite{TSC_graphene_Pangburn1}, containing the normal state Hamiltonian for a single layer graphene and spin-triplet channel in the superconducting pairing as
%~~~~~~~~~~~~~~~~~~~~~~~
\begin{eqnarray}
	\mathcal{H} = H_{\text{SLG}} + H_{\text{triplet}}\ ,
\end{eqnarray}
Where, $H_{\text{SLG}}$ is the normal state Hamiltonian for the monolayer graphene and $H_{\text{triplet}}$ contains the spin-triplet pairings. Those are written as, 
%~~~~~~~~~~~~~~~~~~~~~~~
\begin{eqnarray}
	H_{\text{SLG}} &= -t \sum_{\langle ij \rangle \sigma}  \big( A^{\dagger}_{i\sigma} B_{j\sigma} + H.C. \big) \non \\
	&- \mu \sum_{i \sigma} \big(A^{\dagger}_{i\sigma} A_{i\sigma} + B^{\dagger}_{i\sigma} B_{i\sigma}\big)\ ,
\end{eqnarray}
%~~~~~~~~~~~~~~~~~~~~~~~
\begin{eqnarray}
	H_{\text{triplet}} = H^{x}_{\text{nn}} + H^{y}_{\text{nn}} + H^{z}_{\text{nn}}\ ,
\end{eqnarray}
%~~~~~~~~~~~~~~~~~~~~~~~
With, 
\begin{eqnarray}
	H^{x}_{\text{nn}} = \sum_{\langle ij \rangle} \Delta^{x}_{ij} \big( A^{\dagger}_{i\uparrow} B^{\dagger}_{j\uparrow} - A^{\dagger}_{i\downarrow} B^{\dagger}_{j\downarrow} \big) + H.C. \ ,\\
	H^{y}_{\text{nn}} = \sum_{\langle ij \rangle} \Delta^{y}_{ij} \big( A^{\dagger}_{i\uparrow} B^{\dagger}_{j\uparrow} + A^{\dagger}_{i\downarrow} B^{\dagger}_{j\downarrow} \big) + H.C. \ ,\\
	H^{z}_{\text{nn}} = \sum_{\langle ij \rangle} \Delta^{z}_{ij} \big( A^{\dagger}_{i\uparrow} B^{\dagger}_{j\downarrow} + A^{\dagger}_{i\downarrow} B^{\dagger}_{j\uparrow} \big) + H.C. \ ,
\end{eqnarray}
%~~~~~~~~~~~~~~~~~~~~~~~
where, $A^{\dagger}_{i\sigma}$ represents the creation of electron at sublattice $A$ and at site $i$ with spin $\sigma$. Here, ${\langle ij \rangle}$ represents the nearest neighbors (nn) component. $t$ and $\mu$ represent the nn hopping amplitude and the chemical potential, respectively. Also, $\Delta^{x}_{ij}$, $\Delta^{y}_{ij}$, and $\Delta^{z}_{ij}$ denote the pairing amplitudes of different spin-triplet channels along the nearest neighbors (as shown in Fig.~\ref{mBZ2}(a)).

%~~~~~~~~~~~~~~~~~~~~~~~

By considering the Fourier transformation we translate to the reciprocal space and derive the Hamiltonian as, 
\begin{align}
	H_{\text{SLG}} =& \sum_{k,\sigma} -t \gamma(k) \big( A^{\dagger}_{k\sigma} B_{k\sigma}  + H.C. \big) \non \\
	&- \mu \sum_{k,\sigma} \big( A^{\dagger}_{k\sigma} A_{k\sigma}  + B^{\dagger}_{k\sigma} B_{k\sigma} \big)\ ,\\
	H^{x}_{\text{nn}} =& \sum_{k} f^{x}_{nn}(k) \big( A^{\dagger}_{k\uparrow} B^{\dagger}_{-k\uparrow} - A^{\dagger}_{k\downarrow} B^{\dagger}_{-k\downarrow} \big) + H.C.\ , \\
	H^{y}_{\text{nn}} =& i\sum_{k} f^{y}_{nn}(k) \big( A^{\dagger}_{k\uparrow} B^{\dagger}_{-k\uparrow} + A^{\dagger}_{k\downarrow} B^{\dagger}_{-k\downarrow} \big) + H.C.\ , \\
	H^{z}_{\text{nn}} =& \sum_{k} f^{z}_{nn}(k) \big( A^{\dagger}_{k\uparrow} B^{\dagger}_{-k\downarrow} + A^{\dagger}_{k\downarrow} B^{\dagger}_{-k\uparrow} \big) + H.C.\ ,
\end{align}
with, 
%~~~~~~~~~~~~~~~~~~~~~~~
\begin{eqnarray}
	\gamma(\mathbf{k}) &=& \sum^{3}_{l=1} e^{i \mathbf{k} \cdot \mathbf{\delta}_{l}}, \hspace{5pt}
	f^{x}_{nn}(\mathbf{k}) = \sum^{3}_{l=1} \Delta^{x}_{\delta_{l}} e^{i \mathbf{k} \cdot \mathbf{\delta}_{l}}, \non \\
	f^{y}_{nn}(\mathbf{k}) &=& \sum^{3}_{l=1} \Delta^{y}_{\delta_{l}} e^{i \mathbf{k} \cdot \mathbf{\delta}_{l}}\ .
\end{eqnarray}
%~~~~~~~~~~~~~~~~~~~~~~~
Note that, here $\mathbf{\delta}_{l}$ denotes the nearest neighbour vectors and are given by $\mathbf{\delta}_{1} = (0, -a)$, $\mathbf{\delta}_{2} = (-\sqrt{3}a/2, a/2)$, $\mathbf{\delta}_{3} = (\sqrt{3}a/2, a/2)$. For simplicity, the carbon-carbon distance $a$ is set to unity in the subsequent calculations. On the other hand, depending on the irreducible representation of the crystalline lattice symmetry (here $D_{6h}$ for graphene), one can have non-zero components of the superconducting pairing strengths along the three nn directions. The amplitude of $x$- and $y$- components of the spin-triplet pairing functions along those direction are given by, $\Delta^{x} \equiv \Delta_{\text{sc}} (0, -1/\sqrt{2}, 1/\sqrt{2})^{T}$ and $\Delta^{y} \equiv  \Delta_{\text{sc}} (2/\sqrt{6}, -1/\sqrt{6}, -1/\sqrt{6})^{T}$. Thus we have, 
\begin{eqnarray}
	\gamma(k) &=& e^{-i k_{y}} + 2 e^{\frac{i k_{y}}{2}} \cos\left(\frac{\sqrt{3} k_{x}}{2}\right) \ ,\\
	f^{x}_{nn}(k) &=& i \sqrt{2} \Delta_{\text{sc}} e^{\frac{ik_{y}}{2}} \sin\left(\frac{\sqrt{3} k_{x}}{2}\right)\ , \label{fxnn}\\
	f^{y}_{nn}(k) &=& \frac{2 \Delta_{\text{sc}} }{\sqrt{6}} e^{-i k_{y}} \left[1 - e^{\frac{3 i k_{y}}{2}} \cos\left(\frac{\sqrt{3} k_{x}}{2}\right)\right]
	\label{fynn}\ ,
\end{eqnarray}
Considering these pairing functions $f^{x}_{nn}(k)$ and $f^{y}_{nn}(k)$, along with their low-energy approximations, we construct the low-energy BdG Hamiltonian for tBLG and tDBLG.

%~~~~~~~~~~~~~~~~~~~~~~~~~~~~~~~~~~~~~~~~~~~~~~~~~~~~~~~~~~~~
\section{DERIVATION OF LOW ENERGY BdG HAMILTONIAN FOR GRAPHENE WITH $p_{x} + ip_{y}$ PAIRING}
\label{AppB}
%~~~~~~~~~~~~~~~~~~~~~~~~~~~~~~~~~~~~~~~~~~~~~~~~~~~~~~~~~~~~~

The BdG Hamiltonain for the single layer graphene can be
written in a basis, $\Psi_{\bf{k}} = \bigl( A_{k\uparrow}, B_{k\uparrow}, A_{k\downarrow}, B_{k\downarrow}, A^{\dagger}_{-k\uparrow}, B^{\dagger}_{-k\uparrow}, A^{\dagger}_{-k\downarrow}, B^{\dagger}_{-k\downarrow} \bigr)^{T}$ such that,

\begin{align}
	H_{\text{BdG}}({\bf k})= 
	\left( \begin{array}{cc}
		H_{\text{SLG}}({\bf k}) & H_{\Delta}({\bf k})\\
		H^\dagger_{\Delta}({\bf k}) & -H_{\text{SLG}}({\bf k}) 
	\end{array}\right)\ ,
	\label{Eq:BdG_SLG_TB}
\end{align}	
with, 
\begin{align}
	H_{\text{SLG}}({\bf k}) = 
	\left( \begin{array}{cccc}
		-\mu & -t \gamma ({\bf k}) & 0 & 0\\
		-t \gamma ({\bf k})^{*} & -\mu & 0 & 0\\ 
		0 & 0 & -\mu & -t \gamma ({\bf k})\\
		0 & 0 & -t \gamma ({\bf k})^{*} & -\mu \\ 
	\end{array}\right)\ ,
	\label{Eq:H:BLG1}
\end{align}	
where, $\mu$ is the chemical potential, $t$ denotes the nearest neighbor hopping amplitude and  $\gamma ({\bf k}) =  e^{-ik_{y}} \bigl[ 1 + 2 e^{3 i k_{y}/2} \cos(\frac{\sqrt{3} k_{x}}{2})\bigr]$. Also, the superconducting pairing Hamiltonian can be written as

\begin{align}
	H_{\Delta}({\bf k}) = 
	\left( \begin{array}{cccc}
		0 & f_{+}^{nn}({\bf k}) & 0 & 0\\
		-f_{+}^{nn}({-\bf k}) & 0 & 0 & 0 \\
		0 & 0 & 0 & -f_{-}^{nn}({\bf k})\\
		0 & 0 & f_{-}^{nn}({-\bf k}) & 0 
	\end{array}\right)\ ,
	\label{Eq:H:BLG1}
\end{align}	
where, $f_{+}^{nn} ({\bf k}) = \frac{1}{2} \big(f^{x}_{nn}({\bf k} ) + i f^{y}_{nn}({\bf k}) \big)$, $f_{-}^{nn} ({\bf k}) = \frac{1}{2} \big(f^{x}_{nn}({\bf k} ) - i f^{y}_{nn}({\bf k}) \big)$.

\begin{figure}[h]
	\centering
	\subfigure{\includegraphics[width=0.5\textwidth]{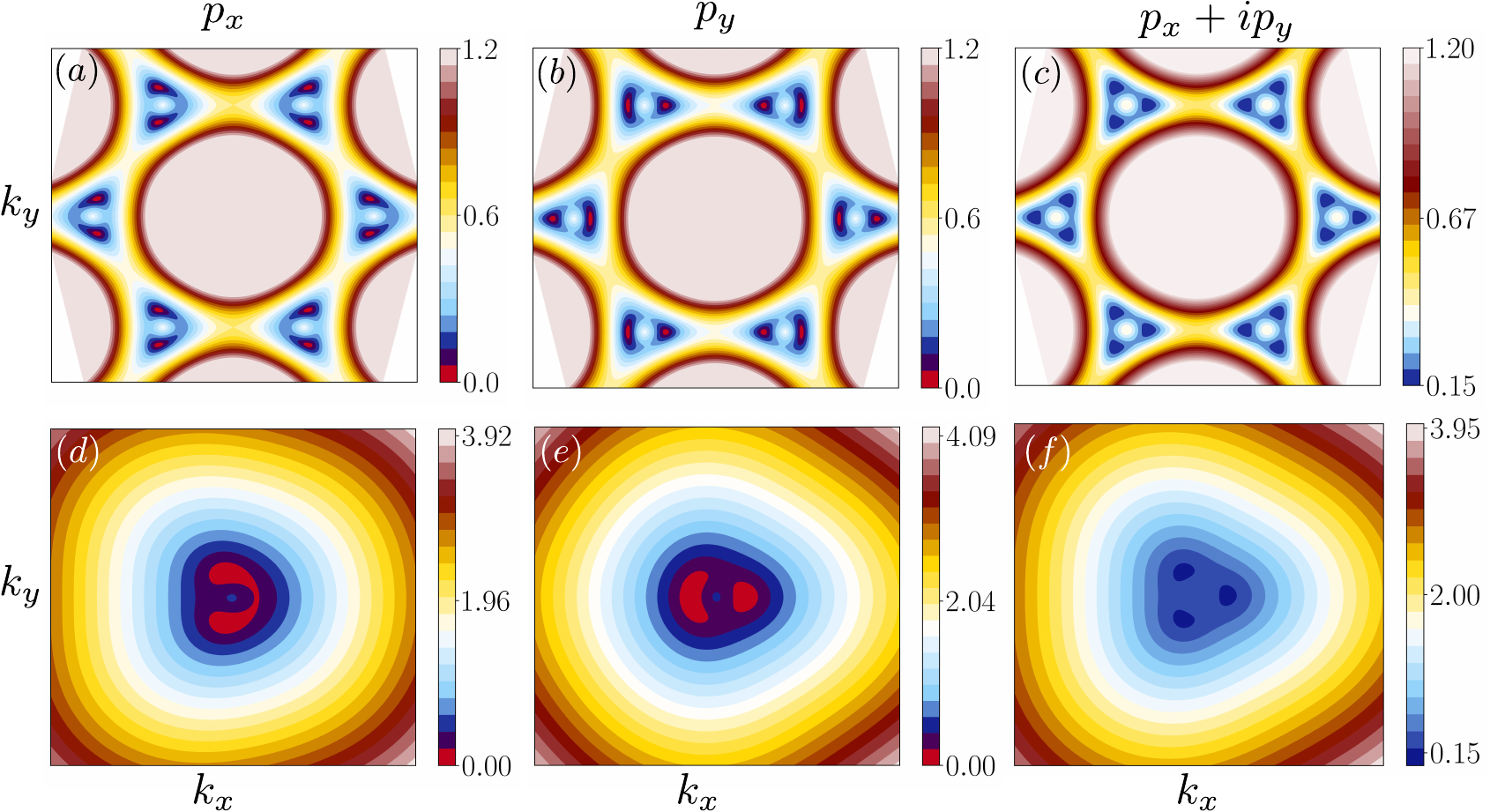}}
	\caption{In panels (a), (b) and (c), lowest energy band ($E>0$) is shown on the BZ respectively for $p_{x}$, $p_{y}$ and $p_{x} + ip_{y}$-superconducting pairings at $\mu = 0.4t$ and $\Delta_{sc} = 0.4t$ considering the tight-binding model. On the other hand, 
	in panels (d), (e) and (f), lowest energy band is shown on the BZ respectively for $p_{x}$, $p_{y}$ and $p_{x} + ip_{y}$-superconducting states  at $\mu = 0.4t$ and $\Delta_{sc} = 0.4t$ implementing the low energy continuum model. Here, we choose $t=1$. 
	}
	\label{pairings}
\end{figure}

Here the form factors $f^{x}_{nn}(\mathbf{k})$ and $f^{y}_{nn}(\mathbf{k})$ for the nearest neighbor $p_{x}$- and $p_{y}$- superconducting pairing for monolayer graphene are given by Eq.~(\ref{fxnn}) and Eq.~(\ref{fynn}). Diagonalizing Eq.~(\ref{Eq:BdG_SLG_TB}), we depict the lowest energy band ($E>0$) 
on the BZ respectively for $p_{x}$, $p_{y}$ and $p_{x} + ip_{y}$-superconducting states in Figs.~\ref{pairings}(a), (b) and (c) choosing $\mu = 0.4t$ and $\Delta_{sc} = 0.4t$. This results reveal the symmetry of the superconducting state and the existence of nodal points in the energy spectrum. From Figs.~\ref{pairings}(a) and (b) we infer that both $p_{x}$- and $p_{y}$-superconducting pairings are nodal and break the $C_{3}$-rotational symmetry around the valley-$K$ and $K^{\prime}$. On the other hand, from Fig.~\ref{pairings}(c) it is evident that $p_{x} + ip_{y}$ superconducting state is gapped and respects the $C_{3}$ rotational symmetry around the Dirac points. 

%------------------------------------------------------------------------------------------
%-------------------------------------------------------------------------------
%------------------------------------------------------------------------------------------
%~~~~~~~~~~~~~~~~~~~~~~~~~~~~~~~~~~~~~~~~~~~~~~~~~~~~~~~~~~~~~~~~~~

To write a low energy continuum model of single layer graphene starting from the BdG Hamiltonian, here we perform Taylor series expansion near valley-$K$ as,
\begin{align}
	t \gamma ({\bf k})|_{K+q} &= -\frac{3 t }{2} (q_{x} + i q_{y})\ ,\\
	f^{x}_{nn}({\bf k})|_{K+q} &=   \frac{i \Delta_{sc} \sqrt{3}}{\sqrt{2}}  \left[ 1 - \frac{1}{2}(q_{x} - i q_{y})\right]\ , \\
	f^{y}_{nn}({\bf k})|_{K+q} &= \frac{\Delta_{sc} \sqrt{3}}{\sqrt{2}}  \left[ 1 + \frac{1}{2}(q_{x} - i q_{y})\right]\ .
\end{align}

The Dirac points of the single layer graphene are located at $K = (\frac{4\pi}{3\sqrt{3}}, 0)$ and $K^{'} = (-\frac{4\pi}{3\sqrt{3}},0)$ in the BZ. Therefore, the BdG Hamiltonian for a single layer of graphene under the low energy approximation can be written as,
\begin{align}
	h_{\text{BdG}}({\bf q})= 
	\left( \begin{array}{cc}
		h_{\text{SLG}}({\bf q}) & h_{\Delta}({\bf q})\\
		h^\dagger_{\Delta}({\bf q}) & -h_{\text{SLG}}({\bf q}) 
	\end{array}\right)\ ,
	\label{BdG_SLG_low}
\end{align}	
where, 
\begin{align}
	h_{\text{SLG}}({\bf q}) = 
	\left( \begin{array}{cccc}
		-\mu & v_{F} q_{+} & 0 & 0\\
		v_{F} q_{-} & -\mu & 0 & 0\\ 
		0 & 0 & -\mu & v_{F} q_{+}\\
		0 & 0 & v_{F} q_{-} & -\mu \\ 
	\end{array}\right)\ ,
	\label{Eq:H:BLG1}
\end{align}	
and
\begin{align}
	h_{\Delta}({\bf q}) = 
	\left( \begin{array}{cccc}
		0 & i \sqrt{\frac{3}{2}} \Delta_{sc} & 0 & 0\\
		-\frac{1}{2} i \sqrt{\frac{3}{2}} \Delta_{sc} q_{+} & 0 & 0 & 0 \\
		0 & 0 & 0 & \frac{1}{2} i \sqrt{\frac{3}{2}} \Delta_{sc} q_{-}\\
		0 & 0 & -i \sqrt{\frac{3}{2}} \Delta_{sc} & 0 
	\end{array}\right)\ .
	\label{SLG_pairing}
\end{align}	

Here, $q_{\pm} = (q_{x} \pm iq_{y})$. Then, we diagonalize Eq.~(\ref{BdG_SLG_low}) and depict the lowest energy band ($E>0$) in the BZ for $p_{x}$, $p_{y}$ and $p_{x} + ip_{y}$-superconducting states in Figs.~\ref{pairings}(d), (e) and (f) respectively 
at $\mu = 0.4t$ and $\Delta_{sc} = 0.4t$. It is observed that all the key features (\ie nodal or gapped, $C_{3}$-rotational symmetry) remain unaltered for the respective superconducting states as mentioned before for the tight-binding model. A close observation further reveals that the nature of the low energy band is similar to the one of the valley of the tight-binding results presented in the upper row of Fig.~(\ref{pairings}).

%~~~~~~~~~~~~~~~~~~~~~~~~~~~~~~~~~~~~~~~~~~~~~~~~~~~~~~~~~~~~~~~~~~~
\section{SPECTRAL FUNCTIONS ANALYSIS FOR tBLG AND tDBLG}
\label{AppC}
%~~~~~~~~~~~~~~~~~~~~~~~~~~~~~~~~~~~~~~~~~~~~~~~~~~~~~~~~~~~~~~~~~~~

%-------------------------------------------------------------------------------
\begin{figure*}[t]
	\centering
	\subfigure{\includegraphics[width=0.8\textwidth]{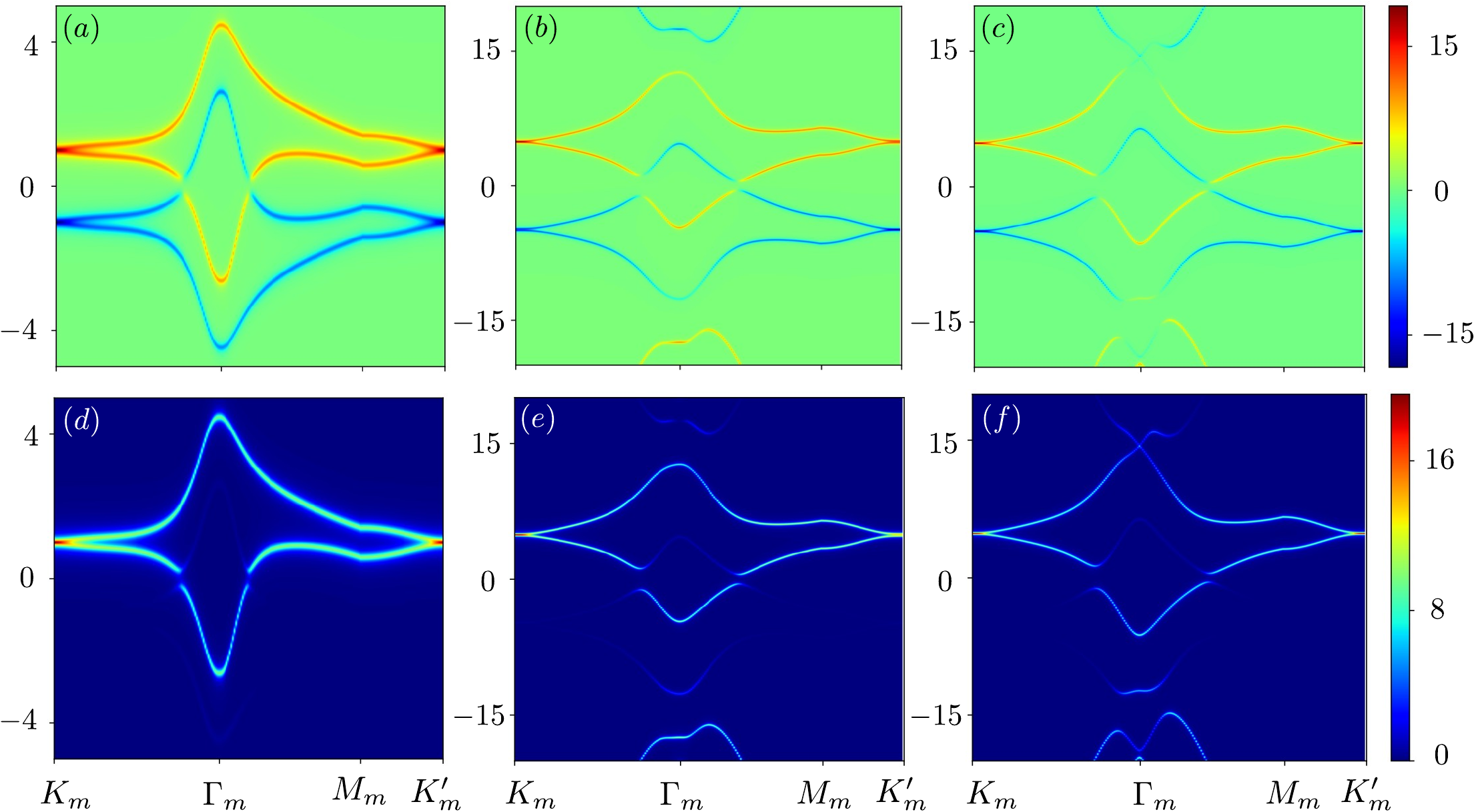}}
	\caption{In the upper panel, we depict the particle-hole resolved spectral functions: in panel (a) for tBLG with $\mu = 1$ meV and $\Delta = 5$ meV, and in panels (b)-(c) for AB-AB and AB-BA stacked tDBLGs respectively with $\mu = 5$ meV and $\Delta = 15$ meV. In the lower panel, we show the electronic spectral functions: in panel (d) for tBLG with $\mu = 1$ meV and $\Delta = 5$ meV, and in panels (e)-(f) for AB-AB and AB-BA stacked tDBLGs respectively choosing $\mu = 5$ meV and $\Delta = 15$ meV.
	}
	\label{spectral_func}
\end{figure*}
%------------------------------------------------------------------------------------------

Here, we calculate both the electronic as well as %spectral function and the 
	particle-hole resolved spectral functions. For that purpose, we define the zero temperature Green's function of the BdG Hamiltonian 
	$H_{\text{BdG}}(\mathbf{q})$ as,
	\begin{align}
		G(\mathbf{q}, \omega) = \big(\omega - H_{\text{BdG}}(\mathbf{q}) + i 0^{+}\big)^{-1}\ ,
	\end{align}
	Thus the electronic spectral function ($A_{el}$) and the hole spectral function ($A_{hl}$) can be written as,
	\begin{align}
		A_{el}(\mathbf{q}, \omega) &= -\frac{1}{\pi} \mathrm{Im} \left[\mathrm{Tr}_{\mathrm{el}}\, G(\mathbf{q}, \omega)\right]\ , \\ 
		A_{hl}(\mathbf{q}, \omega) &= -\frac{1}{\pi} \mathrm{Im} \left[\mathrm{Tr}_{\mathrm{hl}}\, G(\mathbf{q}, \omega)\right]\ ,
	\end{align}
	where, the trace $\mathrm{Tr}_{\mathrm{el}}$ and $\mathrm{Tr}_{\mathrm{hl}}$ is performed respectively over the electron and hole-blocks of the Green's function matrix. To realize the particle-hole character of the BdG quasi-particles we calculate
	\begin{align}
		A_{el-hl}(\mathbf{q}, \omega) &= -\frac{1}{\pi} \mathrm{Im} \mathrm{Tr} \left[\tau_{z} G(\mathbf{q}, \omega)\right]\ ,
	\end{align}
	with $\tau_{z}$ being the Pauli matrix for the particle-hole degrees of freedom. In Fig.~\ref{spectral_func}(a) and Fig.~\ref{spectral_func}(d), we show 
	the particle-hole resolved spectral function and the electronic spectral function respectively for tBLG with $\mu = 1$ meV and $\Delta = 5$ meV. The corresponding band structure is already shown in Fig.~\ref{tBLG_bands}(d) of the main manuscript.
	
The particle-hole resolved spectral function and electronic spectral function respectively for AB-AB tDBLG are shown in 
	Fig.~\ref{spectral_func}(b) and Fig.~\ref{spectral_func}(e), with $\mu = 5$ meV and $\Delta = 15$ meV. Finally, for AB-BA stacked tDBLG, the particle-hole resolved spectral function and the electronic spectral function respectively are depicted in Fig.~\ref{spectral_func}(c) and Fig.~\ref{spectral_func}(f) choosing the same parameter values. The corresponding band structures are already presented in Fig.~\ref{tDBLG_bands}(d) and Fig.~\ref{tDBLG_bands}(h) 
	of the main manuscript, respectively for AB-AB and AB-BA tDBLGs.
	
Note that, all the particle-hole resolved spectral functions corroborate with the corresponding band structures. For all these parameter values, the system remains in a topological superconducting phase. Hence, we observe the band-inversion feature in the particle-hole resolved spectral function.
	
%~~~~~~~~~~~~~~~~~~~~~~~~~~~~~~~~~~~~~~~~~~~~~~~~~~~~~~~~~~~~~~~~~~~~~~~~~~~~~
\section{DESCRIPTION OF THE LOW ENERGY CONTINUUM MODELS}
\label{AppD}
%~~~~~~~~~~~~~~~~~~~~~~~~~~~~~~~~~~~~~~~~~~~~~~~~~~~~~~~~~~~~~~~~~~~~~~~~~~~~~~
\subsection{tBLG}
In literature, tBLG consists of two coupled monolayer graphene sheets that are rotationally misaligned. Here, we briefly outline the construction of our model Hamiltonian following the Bistritzer-MacDonald model~\cite{MacDonald-tBLG,Koshino-tBLG}. 
%~~~~~~~~~~~~~~~~~~~~~~
The primitive lattice vectors for monolayer graphene are given by $\mathbf{a}_{1} = a_{0} (1, 0)$ and $\mathbf{a}_{2} = a_{0} (1/2, \sqrt{3}/2)$, where $a_{0} = 0.246~\mathrm{nm}$ is the lattice constant. The corresponding reciprocal lattice vectors can be written as $\mathbf{b}_{1} = (2\pi / a_{0}) (1, -1 / \sqrt{3})$ and $\mathbf{b}_{2} = (2\pi / a_{0}) (0, 2 / \sqrt{3})$. When the two graphene layers are rotated, their reciprocal lattice vectors rotate by the same angle. We consider that layer-1 and layer-2 are rotated by $-\theta/2$ and $\theta/2$ with respect to each other, and we denote the layers by the index $l$, where $l = 1, 2$ correspond to layer-1 and layer-2, respectively.  After rotation, the reciprocal lattice vectors become $\mathbf{b}^{(l)}_{1,2} = R[(-1)^{l} (\theta / 2)] \mathbf{b}_{1,2}$, where $R(\phi)$ represents the rotation matrix by angle $\phi$. In small-angle twisted systems, a slight mismatch in the lattice periodicity gives rise to a long-period moir\'e unit cell. Consequently, the momentum difference between the rotated layers defines the moir\'e reciprocal lattice vectors as $\mathbf{G}^{m}_{1,2} = \mathbf{b}^{(1)}_{1,2} - \mathbf{b}^{(2)}_{1,2}$ 
(see Fig.~\ref{mBZ2}). After the rotation, the Dirac points corresponding to each monolayer graphene are now located at $K^{(l)}_{\xi} = -\xi (2 \mathbf{b}^{(l)}_{1} + \mathbf{b}^{(l)}_{2})/3$, where $\xi$ is the valley index taking values $\pm 1$ for the valley-$K$ and valley-$K^{\prime}$, respectively.
%~~~~~~~~~~~~~~~~~~~~~~
%tBLG bare model Hamiltonian:
Small-angle twisted systems are best described by an effective low-energy continuum model~\cite{MacDonald-tBLG,Koshino-tBLG}. Here, we write the model Hamiltonian for tBLG in the basis $\Phi_{\bf{q}} = (A_{q_1}, B_{q_1}, A_{q_2}, B_{q_2})^{T}$ as,
\begin{align}
	H^{0,\xi}_{\text{tBLG}}({ \mathbf q})=\left(\begin{array}{cc}%
		h^{\xi}_{0}({\mathbf{q_1}}) & T^{\xi}\\
		{T^{\xi}}^\dagger & h^{\xi}_{0}({\mathbf{q_2}}) 
	\end{array}\right)\ ,
	\label{Eq:tBLG}
\end{align}
with $A_{q_l}$ and $B_{q_l}$ denote the annihilation operators with momentum $q_l$ in layer-$l$ at sublattice-$A$ and $B$, respectively. Also, $h^{\xi}_{0}({\mathbf{q_1}})$ and $h^{\xi}_{0}({\mathbf{q_2}})$ are Hamiltonians for two monolayers of tBLG with $\mathbf{q}_{(l)} = R[(-1)^{(l)} (\theta/2)]\mathbf{q}$ near valley-$K_\xi$. Here, $T^{\xi}$ describes the interlayer coupling between the two layers. 
%~~~~~~~~~~~~~~~~~~~~~~
The low energy model Hamiltonian for a single layer of graphene near valley-$K_\xi$ is given by,
\begin{align} 
	h^{\xi}_{0}({ \mathbf q}) &=
	\left( \begin{array}{cc}
		-\mu  & \hbar v_{F} q^{\xi}_+ \\
		\hbar v_{F} q^{\xi}_- & -\mu
	\end{array}\right)\ ,
	\label{Eq:low_SLG}
\end{align}
where, $q_{\pm} = \xi q_{x} \pm i q_{y}$ and $\mu$ is the chemical potential. In our calculations, we  consider, $ \hbar v_{F}/a_{0} =2135.4$ meV~\cite{RMP-Graphene-review,McCann_2013-BilayerReview,Koshino-tBLG}.
%~~~~~~~~~~~~~~~~~~~~~~
The effective interlayer coupling between the two layers (\ie tBLG) is given by the matrix $T$ as follows~\cite{MacDonald-tBLG,Koshino-tBLG},
\begin{align}
	T^{\xi} = T_{1,\xi} + T_{2,\xi} e^{i\xi \mathbf{G_{1}^{m}} \cdot \mathbf{r}} + T_{3,\xi} e^{i\xi (\mathbf{G_{1}^{m}} + \mathbf{G_{2}^{m}}) \cdot \mathbf{r}}\ ,
	\label{Eq:H:AB-AB-Full1}
\end{align}
where, $\mathbf{G^{m}} = n_{1}\mathbf{G_{1}^{m}} + n_{2}\mathbf{G_{2}^{m}}$ is the reciprocal lattice vector in the mBZ (the superscript $m$ in $\mathbf{G}$'s denotes that) and $n_{1}$, $n_{2}$ are integers. The tunneling matrices in Eq.~(\ref{Eq:H:AB-AB-Full1}) are given by
\begin{align}
	T_{1,\xi} &=
	\left( \begin{array}{cc}
		u  & u^{\prime} \\
		u^{\prime} & u
	\end{array}\right) ~~,~~
	T_{2,\xi} 
	&=
	\left( \begin{array}{cc}
		u  & u^{\prime} \omega^{-\xi} \\
		u^{\prime}\omega^{\xi} & u
	\end{array}\right) ~~,~~ \nonumber\\
	&T_{3,\xi} 
	=&\!\!\!\!\!\!\!\!\!\!\!\!\!\!\!\!\!\!\!\!
	\left( \begin{array}{cc}
		u  & u^{\prime} \omega^{\xi} \\
		u^{\prime}\omega^{-\xi} & u
	\end{array}\right)\ .
	\label{Eq:H:BLG:1}
\end{align}
Here, $u$ and $u^{\prime}$ stand for tunneling amplitudes between AA/BB and AB/BA sublattices respectively. We choose $u = 79.7$ meV, $u^{\prime} = 97.5$ meV~\cite{Koshino-tBLG,corrugation-DFT-uu,Dai2016-uu} and $\omega = e^{2 \pi i/3}$. 
%~~~~~~~~~~~~~~~~~~~~~~
%-------------------------------------------------------------------------------
%------------------------------------------------------------------------------
\begin{figure*}[t]
	\centering
	\subfigure{\includegraphics[width=0.8\textwidth]{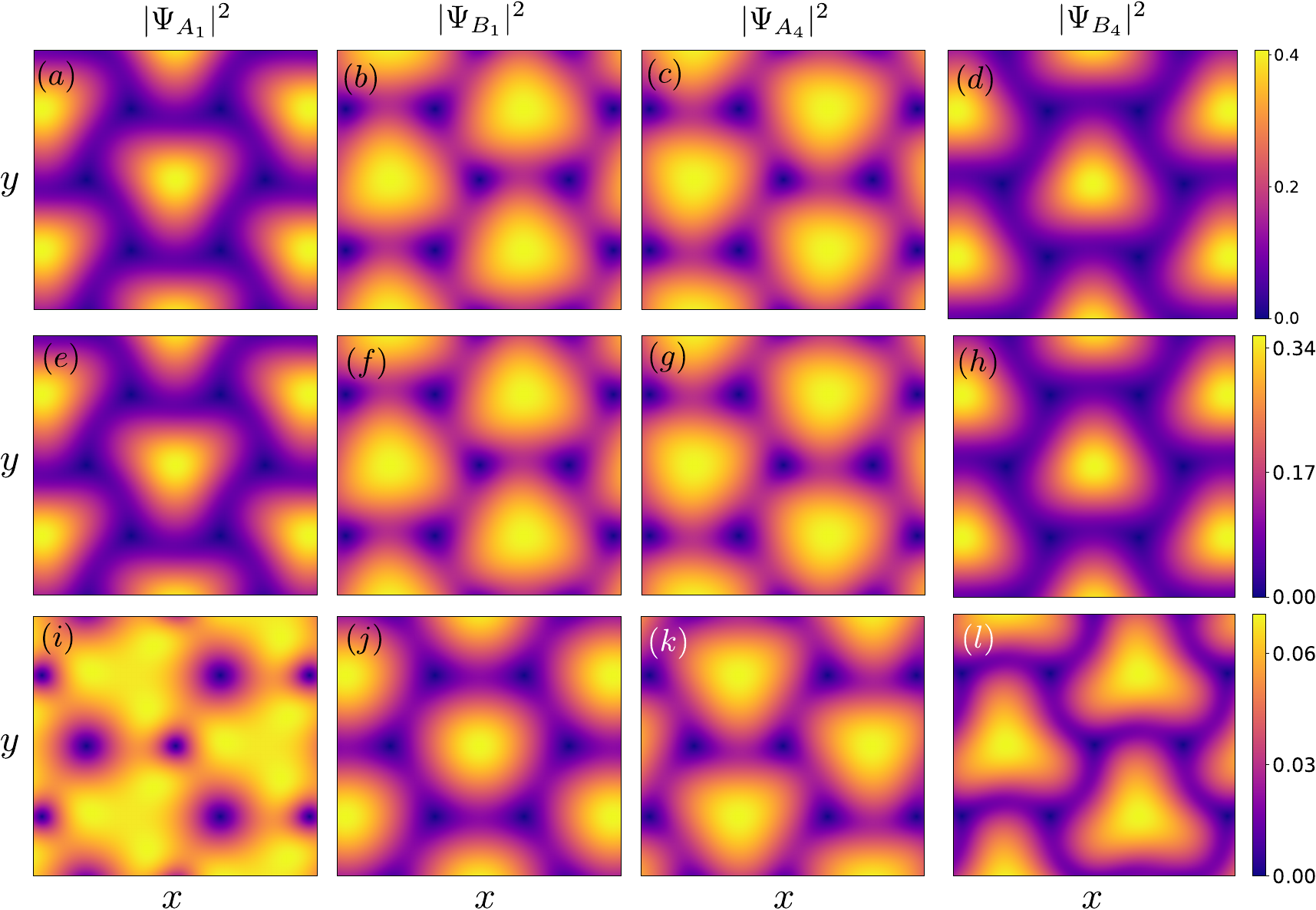}}
	\caption{(a) Density plot for the squared amplitude of Bloch states are depicted in the $x$ - $y$ plane for AB-AB tDBLG at a twist angle $\theta = 1.3^{o}$ near valley-$K$ and with the spin-up $p_{x} + ip_{y}$ superconducting order. These states belong to the $\Gamma_{m}$-point of the first conduction band. The four columns respectively refer to the sublattice-A of layer-1 ($|\psi_{A_{1}}|^{2}$), sublattice-B of layer-1 ($|\psi_{B_{1}}|^{2}$), sublattice-A of layer-4 ($|\psi_{A_{4}}|^{2}$) and sublattice-B of layer-4 ($|\psi_{B_{4}}|^{2}$). The first row (\ie panel (a) - (d)), second row (i.e., panel (e) - (h)) and third row (i.e., panel (i) - (l)) correspond to $\mu = 0.5$ meV, $\mu = 5$ meV, $\mu = 15$ meV respectively, choosing $\Delta_{\mathrm{sc}} = 10$ meV 
	in case of AB-AB tDBLG.}
	\label{tDBLG_Bloch_L14}
\end{figure*}
%------------------------------------------------------------------------------------------
%------------------------------------------------------------------------------------------
%~~~~~~~~~~~~~~~~~~~~~~
\subsection{tDBLG}
%~~~~~~~~~~~~~~~~~~~~~
When two bilayer graphene sheets are stacked on top of each other and slightly rotated, it forms a tDBLG. However, depending on the type of stacking of the bilayers, the double bilayer exhibits two stable configurations: one is AB-AB and the other one is AB-BA (see Fig.~\ref{mBZ2}(b)). Below we write the low energy continuum model for both types of tDBLG near valley-$K_{\xi}$ in the basis $\chi_{\bf{q}} = (A^{1}_{q_1}, B^{1}_{q_1}, A^{2}_{q_1}, B^{2}_{q_1}, A^{3}_{q_2}, B^{3}_{q_2}, A^{4}_{q_2}, B^{4}_{q_2})^{T}$ as,
\begin{align}
	H^{0,\xi}_{\text{AB-AB}}=\left(\begin{array}{cc}%
		H^{0,\xi}_{AB}({\mathbf{q_1}}) & \tilde{T}^{\xi}\\
		(\tilde{T}^{\xi})^\dagger & H^{0,\xi}_{AB}({\mathbf{q_2}}) 
	\end{array}\right)\ ,
	\label{Eq:ABAB_tDBLG}
\end{align}
%~~~~~~~~~~~~~~~~~~~~~~
\begin{align}
	H^{0,\xi}_{\text{AB-BA}}=\left(\begin{array}{cc}%
		H^{0,\xi}_{AB}({\mathbf{q_1}}) & \tilde{T}^{\xi}\\
		(\tilde{T}^{\xi})^\dagger & H^{0,\xi}_{BA}({\mathbf{q_2}}) 
	\end{array}\right)\ ,
	\label{Eq:ABBA_tDBLG}
\end{align}
where, $A^{j}_{q_l}$ and $B^{j}_{q_l}$ correspond to the annihilation operators with momentum $q_l$ in layer-$j$ and at sublattice-$A$ and $B$, respectively. Here, $q_1$ and $q_2$ belong to the rotated BZ of first and second bilayer graphene respectively. Further, $H^{0,\xi}_{AB}({\mathbf{q}})$ and $H^{0,\xi}_{BA}({\mathbf{q}})$ represent the Hamiltonians for two bilayers of tDBLG with AB and BA stacking configurations respectively. Here, we assume that 
the first bilayer is rotated by $-\theta/2$ and the second bilayer follows a rotation of $\theta/2$ with respect to each other.  
%~~~~~~~~~~~~~~~~~~~~~~
The low energy continuum Hamiltonian for a AB-stacked and BA-stacked bilayer graphene near valley-$K_{\xi}$ in the basis $(A^{1}_{q},B^{1}_{q},A^{2}_{q},B^{2}_{q})$ can be written as,
\begin{align}
	H^{0}_{\rm{AB}}({ \bf q}) = 
	\left( \begin{array}{cc}
		h_{0}({ \bf q}) & g^\dagger ({ \bf q})\\
		g({ \bf q}) & h_{0}({ \bf q}) \end{array}\right)\ ,~
	H^{0}_{\rm{BA}}({ \bf q}) = 
	\left( \begin{array}{cc}
		h_{0}({ \bf q}) & g({ \bf q})\\
		g^\dagger({ \bf q}) & h_{0}({ \bf q}) \end{array}\right)\ .
	\label{Eq:H:BLG1}
\end{align}	
where, $h_{0}(\bf q)$ denotes the low energy model Hamiltonian for the monolayer graphene near valley-$K_{\xi}$ (see Eq.~(\ref{Eq:low_SLG})). Here, $g({ \bf q})$ denotes the inter-layer coupling between the two layers of the tDBG and is given by,
\begin{align} 
	g({\bf q}) 
	=
	\left( \begin{array}{cc}
		0  & \gamma_1 \\
		\hbar v_{3} q_{+}  & 0
	\end{array}\right)\ .
	\label{Eq:BLG_interlayer}
\end{align}	
%~~~~~~~~~~~~~~~~~~~~~~
Here, $\gamma_1$ in $g({\bf q})$ (see Eq.~(\ref{Eq:BLG_interlayer})) provides the coupling between the dimer sites (\ie sites in bilayer graphene those are exactly on top of each other). The parameter $v_{3}$ accounts for the coupling between the non-dimer sites of the two layers and generates the trigonal warping of the bilayer graphene Fermi surface. In our calculation we consider the following values of the parameter set $\gamma$'s, $\gamma_1 = 400$ meV meV, $\gamma_3 = 320$ meV~\cite{McCann_2013-BilayerReview,Koshino-tBLG,McCann_2011-BilayerReview}. Also, $\gamma_{3}$ is related to $v_{3}$ as follows, $v_{3} = \sqrt{3} \gamma_{3} a_{0}/2 \hbar$.
To discuss our results in the latter text, we consider both the situation \ie without (minimal model of tDBLG) and with trigonal warping.
%~~~~~~~~~~~~~~~~~~~~~~
The effect of twist appears in between the layer-2 and layer-3 of the four layers of tDBLG. As the BZ of the bilayer graphene is same as the monolayer graphene, the rotation between the two bilayers create the mBZ as we have encountered for tBLG. However, now those interlayer coupling matrices are slightly different than tBLG and are given by~\cite{Koshino-tDBLG},
\begin{align}
	\tilde{T_{i}}^{\xi} &=
	\begin{pmatrix}
		0  & 0 \\
		1 & 0 \\
	\end{pmatrix} \otimes T_i^{\xi}\ .
	\label{Eq:H:BLG:2}
\end{align}

We note that out-of-plane corrugation effects remain relevant for tDBLG too, for the two layers that participate directly in the twist mechanism.
Several early works have incorporated these corrugation effects in modeling tDBLG~\cite{Koshino-tDBLG,Lin_tDBLG}. Hence, we also consider the same parameter values for tunneling amplitudes $u$ and $u^{\prime}$ between the sublattices, as considered for tBLG. Moreover, in our numerical calculations, for both moir\'e systems (\ie tBLG and tDBLGs), the momentum cutoff is chosen to be $4|\mathbf{G}^{m}_{1,2}|$ (as $|\mathbf{G}^{m}_{1,2}|$ is defined before) in the momentum space lattice of mBZ. Furthermore, unlike tBLG in which the magic angle is $1.05^{\circ}$, tDBLG does not exhibit a single twist angle at which the renormalized Fermi velocity vanishes. Instead, previous studies suggest that within a narrow range of twist angles near $\theta = 1.3^{\circ}$, the bands attain maximal flatness, and we consider that as the magic angle for tDBLG in our study~\cite{Haddadi-tDBLG}.

%~~~~~~~~~~~~~~~~~~~~~~~~~~~~~~~~~~~~~~~~~~~~~~~~~~~~~~~~~~~~~~~~~~~~~~~~~~~~~~~~~~~~
\section{BLOCH STATE AMPLITUDE PROFILES FOR OUTER LAYERS OF TDBLG}
\label{AppE}
%~~~~~~~~~~~~~~~~~~~~~~~~~~~~~~~~~~~~~~~~~~~~~~~~~~~~~~~~~~~~~~~~~~~~~~~~~~~~~~~~~~~~
Here, in Fig.~\ref{tDBLG_Bloch_L14}, we present the spatial profiles for the first and fourth (outer) layers of AB-AB tDBLG, using the same parameter values as mentioned in the main text for second and third layers. The four columns correspond to Bloch state amplitude profiles for sublattice-A 
of layer-1 ($|\psi_{A_{1}}|^{2}$), sublattice-B of layer-1 ($|\psi_{B_{1}}|^{2}$), sublattice-A of layer-4 ($|\psi_{A_{4}}|^{2}$), and sublattice-B of layer-4 ($|\psi_{B_{4}}|^{2}$), respectively. The first row (panels (a)-(d)), second row (panels (e)-(h)), and third row (panels (i)-(l)) correspond to $\mu = 0.5~\mathrm{meV}$, $\mu = 5~\mathrm{meV}$, and $\mu = 15~\mathrm{meV}$, respectively, for a fixed pairing amplitude $\Delta_{\mathrm{sc}} = 10~\mathrm{meV}$ 
in case of AB-AB tDBLG. Although the individual spatial profiles differ from those of layers 2 and 3, the qualitative distinction associated with the change in the topological phase remains non-conclusive.

\bibliography{bibfile}{}

%============End of MAIN PAPER=============

\end{document}